\documentclass[preprint]{aastex}

\def\Sec{\hbox{${}^{\prime\prime}$\llap{.}}}
\defcitealias{Cot2007}{C07} 	  	 
\defcitealias{Lau2007}{L07} 	  	 
\defcitealias{Geb1996}{G96}
\defcitealias{Kor2009}{K09}

\usepackage{natbib}

\shorttitle{Central Deprojected Profiles of Early-Type Galaxies}
\shortauthors{Glass et al.}

\begin{document}

\title{The ACS Fornax Cluster Survey. IV. Deprojection of the Surface Brightness Profiles of Early-Type Galaxies in the Virgo and Fornax Clusters: Investigating the ``Core/Power-Law Dichotomy''}

\author{Lisa Glass\altaffilmark{1,2}, Laura Ferrarese\altaffilmark{2}, Patrick C{\^o}t{\'e}\altaffilmark{2}, Andr{\'e}s Jord{\'a}n\altaffilmark{3,4}, Eric Peng\altaffilmark{5, 6}, John P. Blakeslee\altaffilmark{2}, Chin-Wei Chen\altaffilmark{2,7}, Leopoldo Infante\altaffilmark{3}, Simona Mei\altaffilmark{8,9}, John L. Tonry\altaffilmark{10}, \& Michael J. West\altaffilmark{11}}

\altaffiltext{1}{Department of Physics and Astronomy, University of Victoria, Victoria, BC, V8W 3P6, Canada}
\altaffiltext{2}{Herzberg Institute of Astrophysics, National ResearchCouncil of Canada, Victoria, BC, V9E 2E7, Canada}

\altaffiltext{3}{Departamento de Astronom\'ia y Astrof\'isica, Pontificia Universidad Cat\'olica de Chile, Av. Vicu\~na Mackenna 4860, Macul 7820436, Santiago, Chile}

\altaffiltext{4}{Harvard-Smithsonian Center for Astrophysics, 60 Garden St., Cambridge, MA, 02138, USA}

\altaffiltext{5}{Department of Astronomy, Peking University, Beijing 100871, China}

\altaffiltext{6}{Kavli Institute for Astronomy and Astrophysics, Peking University, Beijing 100871, China}

\altaffiltext{7}{Institute of Astronomy, National Central University, Jhongli 32054, Taiwan}

\altaffiltext{8}{University of Paris Denis Diderot,  75205 Paris Cedex 13, France}

\altaffiltext{9}{GEPI, Observatoire de Paris, Section de Meudon, 5 Place J. Janssen, 92195 Meudon Cedex, France}

\altaffiltext{10}{Institute for Astronomy, University of Hawaii, 2680 Woodlawn Drive, Honolulu, HI 96822, USA}

\altaffiltext{11}{European Southern Observatory, Alonso de Cordova 3107, Vitacura, Santiago, Chile}

\begin{abstract}

Although early observations with the \emph{Hubble Space Telescope} (\emph{HST}) pointed to a sharp dichotomy among early-type galaxies in terms of the logarithmic slope $\gamma'$ of their central surface brightness profiles, several studies in the past few years have called this finding into question. In particular, recent imaging surveys of $143$ early-type galaxies belonging to the Virgo and Fornax Clusters using the Advanced Camera for Surveys (ACS) on board \emph{HST} have not found a dichotomy in $\gamma'$, but instead a systematic progression from central luminosity deficit to excess relative to the inward extrapolation of the best-fitting global S\'ersic model. Given that earlier studies also found that the dichotomy persisted when analyzing the deprojected density profile slopes, we investigate the distribution of the three-dimensional luminosity density profiles of the ACS Virgo and Fornax Cluster Survey galaxies. Having fitted the surface brightness profiles with modified S\'{e}rsic models, we then deproject the galaxies using an Abel integral and measure the inner slopes $\gamma_{\rm 3D}$ of the resulting luminosity density profiles at various fractions of the effective radius $R_e$. We find no evidence of a dichotomy, but rather, a continuous variation in the central luminosity profiles as a function of galaxy magnitude. We introduce a parameter, $\Delta_{\rm 3D}$, that measures the central deviation of the deprojected luminosity profiles from the global S\'{e}rsic fit, showing that this parameter varies smoothly and systematically along the luminosity function.

\end{abstract}


\section{Introduction}
\label{sec: Introduction}

The launch of the \emph{Hubble Space Telescope} (\emph{HST}) two decades ago made it possible to study the innermost regions of galaxies at spatial resolutions that were previously unattainable at optical wavelengths. The first \emph{HST} imaging surveys of bright early-type galaxies agreed in finding a luminosity-dependent structural dichotomy in the central brightness profiles --- within the innermost few hundred parsecs, galaxies brighter than $M_B\sim-20.5$~mag showed surface brightness profiles that increased very gently towards the center (``core" galaxies) while fainter galaxies exhibited steeper surface brightness cusps (``power-law" galaxies; e.g., \citealt{Fer1994, Lau1995, Fab1997}). The paucity of galaxies with intermediate slopes was striking --- in plots of radially-scaled luminosity density profiles, core and power-law galaxies were seen to define two distinct, and virtually non-overlapping, populations (see, e.g., Figure 3 of \citealt{Geb1996}; hereafter \citetalias{Geb1996}).

However, as subsequent studies targeted larger and better defined samples, some galaxies having intermediate slopes were discovered.  The distinction between core and power-law galaxies became either less pronounced  (e.g., \citealt{Res2001, Rav2001, Lau2007}, hereafter \citetalias{Lau2007}) or disappeared entirely (\citealt{Fer2006astroph, Fer2006, Cot2007}; hereafter \citetalias{Cot2007}. Note that these later studies parameterized the surface brightness profiles using modified S\'{e}rsic profiles rather than the so-called ``Nuker'' profiles used by earlier authors. See \S3.3.1 of \citealt{Fer2006} for a more detailed discussion.) 

In particular, \citetalias{Cot2007} utilized high-quality \emph{HST} imaging from the Advanced Camera for Surveys Virgo and Fornax Cluster Surveys (ACSVCS, \citealt{Cot2004}; ACSFCS, \citealt{Jor2007}). Taken together, these two surveys represent the largest and most homogeneous imaging database currently available for a well characterized sample of early-type galaxies located in low-mass galaxies clusters in the local universe (i.e., at distances $d \lesssim 20$ Mpc). The distribution of surface brightness profiles for the $\sim140$ ACSVCS/FCS galaxies was found to be a smoothly varying function of galaxy magnitude: galaxies brighter than $M_B \sim -20$~mag showed central luminosity ``deficits"  (typically within $\sim40-200$ pc) with respect to the inward extrapolation of the S\'{e}rsic model that best fit the outer parts of the profiles, gradually transitioning toward the fainter galaxies that showed central luminosity ``excesses" with respect to the S\'ersic law (\citealt{Cot2006, Fer2006}). \citetalias{Cot2007} further showed that a bimodality in the central slopes could be introduced by using a biased sample: in particular, Monte-Carlo simulations showed that the bimodal {\it luminosity} distribution of galaxies observed by \citetalias{Lau2007} would lead naturally to a bimodal {\it slope} distribution, even when the intrinsic slope distribution was continuous along the galaxy luminosity function. 

Since \citetalias{Cot2007} was published, Kormendy~et~al.~(2009; hereafter \citetalias{Kor2009}) have commented on the core/power-law dichotomy issue as well, although they did not compute inner profile slopes. They extracted surface brightness profiles from  40 of the 100 ACSVCS galaxies and combined them with profiles from other space- and ground-based photometry, in some cases adding somewhat to the radial extent of the data. They also included profiles from space- and ground-based imaging of three additional galaxies, NGC~4261, NGC~4636, and M32. Their fits to the surface brightness profiles were determined in a very similar manner to \citetalias{Cot2007}, i.e., fitting modified S\'{e}rsic models (see \S2.1 of \citetalias{Cot2007} and Appendix~A of \citetalias{Kor2009}) and, as such, there were no systematic differences in the fits for individual galaxies, as shown in Figure~75 of \citetalias{Kor2009}. In fact, \citetalias{Kor2009} confirmed the trend from central light deficit to excess along the luminosity function of this sample that was noted by \citet{Fer2006} and \citet{Cot2006, Cot2007}. However, \citetalias{Kor2009} excluded 60\% of the ACSVCS sample -- in particular, the vast majority of galaxies in the $-21.5 \la M_B \la -18.5$ range -- and, unlike \citetalias{Cot2007}, included none of the Fornax cluster galaxies. They consequently found a qualitative gap in the inner slopes of their surface brightness profiles (see their Figure~ 40) and interpreted this gap as confirming the existence of the core/power-law dichotomy. P.~C{\^o}t{\'e}~et~al.~(2011, \emph{in prep}) will provide a much more thorough comparison of the ACSVCS/FCS results with \citetalias{Kor2009}.

Several previous authors (e.g., \citetalias{Geb1996}; \citetalias{Lau2007}) who claimed a dichotomy in central surface brightness slopes, extended their work by examining the slopes of three-dimensional (i.e., deprojected) luminosity density profiles. These studies again found that a dichotomy exists, a result that cannot be immediately assumed given how rapidly shallow projected inner profiles deproject to relatively steeper inner profiles (see, e.g., \citealt{Deh1993, Mer1996}; \citetalias{Geb1996}; Figure~\ref{fig:gzcompare}\emph{a-c} of this paper). To address this issue, we show here that the distribution of slopes noted by \citet{Fer2006astroph, Fer2006} and \citetalias{Cot2007} remains continuous once the profiles are deprojected into three-dimensional luminosity density profiles. The deprojections --- which are based on a numerical inversion of the parameterized surface brightness profiles under the assumption of sphericity --- produce individual inner slopes that are consistent with those obtained using the non-parametric methodologies of \citetalias{Geb1996} and \citetalias{Lau2007}. This finding provides additional support for the conclusion that the apparent division of galaxies into core and power-law types is a consequence of the galaxy selection function used in previous studies, which was greatly overabundant in luminous core galaxies, while galaxies in the magnitude range corresponding to the transition between core and power-law types were under-represented.  (See Figure 4 of \citetalias{Cot2007}.) At the same time, we note that the characterization of galaxies by the slopes of the central brightness profiles is rather sensitive to a number of factors (including the choice of measurement radius, resolution, and model parameterization). We introduce a parameter, $\Delta_{\rm 3D}$, that quantifies the central deviation of the luminosity density profile from the inward extrapolation of the S\'{e}rsic model fitted to the main body of the galaxy. We show that, when parameterized in this way, early-type galaxies show a systematic progression from central luminosity deficits to excesses (nuclei) along the luminosity function.

\section{Observations}

\subsection{The ACS Virgo and Fornax Cluster Surveys}

The ACS Virgo and Fornax Cluster Surveys imaged 143 early-type galaxies (morphological types: E, S0, dE, dE,N, dS0, and dS0,N) using the ACS Wide Field Channel (WFC) in the F475W ($g$) and F850LP ($z$) filters. For the purpose of this work, five galaxies were excluded because of severe dust obscuration in the inner region. The galaxies used in this analysis are listed in Tables~\ref{tab:acsvcs} and \ref{tab:acsfcs}. Combined, the survey galaxies span a $B$-band luminosity range of $\sim750$. The ACSVCS is magnitude-limited down to $B  \approx 12$~mag ($M_B \approx -19.2$~mag, i.e. $\sim$ 1-1.5~mag fainter than the expected core/power-law transition) and $\sim50\%$ complete down to its limiting magnitude of $B \approx 16$~mag ($M_B \approx -15.2$~mag). The ACSFCS sample is complete down to its limiting magnitude of $B \approx 15.5$~mag ($M_B \approx -16.1$~mag). 

The ACS/WFC consists of two $2048 \times 4096$ pixel CCDs with a spatial scale of $0\Sec05$ per pixel, covering a field of view of roughly $202'' \times 202''$. The $0\Sec1$ spatial resolution corresponds to $\approx8.0$~pc in Virgo ($d\approx16.5$~Mpc; \citealt{Mei2007}) and $\approx9.7$~pc in Fornax ($d\approx20.0$~Mpc; \citealt{Bla2009}).
Azimuthally averaged surface brightness profiles were generated for each galaxy, in each band, as explained in detail in \citet{Fer2006} and \citet{Cot2006}. These papers provide full details on corrections for dust obscuration, masking of background sources, the identification of offset nuclei via centroid shifts, and the weighting schemes and minimization routines used in fitting the one-dimensional profiles.
	
\subsection{Parameterization of the Surface Brightness Profiles}
\label{sec: Parameterization}
To deproject the brightness profiles for the program galaxies, point-spread function (PSF) convolved parametric models were fitted to the observed surface brightness profiles derived from both the F475W ($g$-band) and F850LP ($z$-band) images. Parametric models represent the profiles before PSF convolution; in what follows, when discussing a comparison between  ``model'' and ``observed'' profiles, it will be implicitly assumed that the model is first PSF convolved.   

\begin{figure}
\begin{center}
\includegraphics[width=5 in]{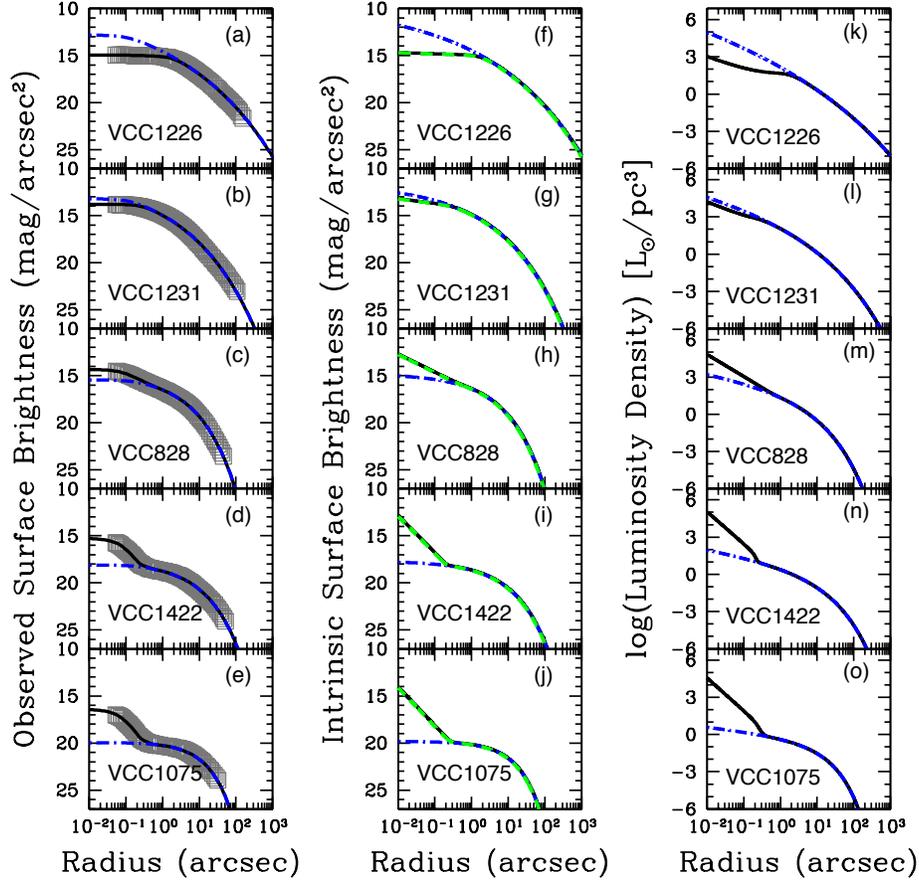}
\caption{Observed $z$-band surface brightness profiles ({\it left panels}), intrinsic (i.e., not PSF-convolved) surface brightness profiles ({\it middle panels}), and luminosity density profiles ({\it right panels}) for five representative galaxies from the ACSVCS: VCC~1226 (= M49 = NGC~4472, with $M_B \approx -21.9$~mag and $R_b \approx 1\Sec8 \approx 142$~pc),  VCC~1231  ($M_B \approx - 19.9$~mag and $R_b \approx 0\Sec3 \approx 20$~pc), VCC~828 ($M_B \approx -18.6$~mag and $R_b \approx 0\Sec5 \approx 43$~pc), VCC~1422 ($M_B \approx -17.4$~mag and $R_b \approx 0\Sec2 \approx 16$~pc), and VCC~1075 ($M_B \approx -16.1$~mag and $R_b \approx 0\Sec3 \approx 21$~pc). In the left column of panels, the gray squares (appearing as a thick line) are the observed surface brightness profiles, the black lines are the PSF-convolved best-fit profiles, and the blue dot-dashed lines indicate the underlying galaxies, i.e., the S\'{e}rsic component by itself. In the middle column of panels, the 
intrinsic models (i.e., without PSF convolution) are shown, with the same colour scheme as the panels on the left. The green dashed lines in the middle panels show the integration of the luminosity density profiles (black lines in the right column of panels) along the line of sight as a test to ensure they reproduce the surface brightness profiles (which they do). The deprojections of the S\'{e}rsic components are shown as blue dot-dashed lines in the rightmost panels.}
\label{fig:profiles}
\end{center}
\end{figure}

Over the vast majority of their radial ranges, the  global brightness profiles of the galaxies in our sample are well represented by (PSF-convolved) S\'{e}rsic $r^{1/n}$ models \citep{Ser1963, Ser1968}, however, in the innermost regions -- typically within $2.0^{+2.5}_{-1.0}\%$ of the effective radius $R_e$ (\citetalias{Cot2007}) -- the surface brightness profiles tend to diverge from a S\'{e}rsic model. For galaxies brighter than $M_B\sim-20$~mag, the surface brightness profiles within $\sim2\%R_e$ fall \emph{below} the global best-fit S\'{e}rsic models. For galaxies with $-20 \lesssim M_B \lesssim -19.5$~mag, a single S\'{e}rsic model generally provides an acceptable fit over all radii, including the innermost regions. Galaxies fainter than $M_B\sim-19.5$~mag tend to have surface brightness profiles that, within $\sim2\%R_e$, extend significantly \emph{above} the inward extrapolation of the global S\'{e}rsic models. In what follows, we shall refer to these light excesses as ``stellar nuclei'', or simply ``nuclei'' (consistent with \citealt{Fer2006}).\footnote{A note on terminology: whereas these bright regions at the centers of early-type galaxies have historically been referred to as ``stellar nuclei'' and the host galaxies as ``nucleated'', groups studying what are likely the same type of objects at a different point in their evolution in late-type galaxies tend to refer to them as ``nuclear star clusters'' or simply ``nuclear clusters'' (e.g., \citealt{Ros2006, Bok2007}). In early types, they have also been referred to as ``light excesses'' (e.g., \citealt{Cot2006}, \citetalias{Cot2007}) or ``extra light'' (e.g., \citealt{Kor1999}; \citealt{Kor2009}). For the sake of simplicity (and because some nuclei --- such as in VCC~1146 --- are in fact disk-like structures, therefore making the term ``cluster'' somewhat misleading in these cases) we refer to them here as ``compact stellar nuclei'' in ``nucleated'' galaxies. In any case, the practical definition of a nucleus is the same as that adopted in all previous papers in this and the ACSVCS series: i.e., ``a central excess in the brightness profile relative to the fitted S\'{e}rsic model'' (i.e., Appendix A of \citealt{Cot2006}).} As one moves down the galaxy luminosity function --- and the surface brightness of the underlying galaxy drops in kind --- these nuclei become increasingly obvious in both the \emph{HST} images and the 1D surface brightness profiles (see, e.g., Figures 1 and 2 of \citetalias{Cot2007}, as well as our Figure~\ref{fig:profiles}\emph{a-e}).

We can account for central luminosity variations from a global S\'{e}rsic fit using a single fitting function, the ``core-S\'{e}rsic'' model (e.g., \citealt{Gra2003}; \citealt{Tru2004}), in which the S\'{e}rsic model is modified to have a power-law profile inside a break radius, $R_b$:
\begin{eqnarray}
\label{eq: core-Sersic}
I(R) &=& I^{\prime} \left[ 1 + \left( \frac{R_b}{R} \right) ^{\alpha} \right] ^{\gamma/\alpha}\times \exp \left[ -b_n \left( \frac{R^{\alpha} + R^{\alpha}_b} {R^{\alpha}_e}\right) ^{1/(\alpha n)} \right],
\end{eqnarray}
where $I^{\prime}$ is related to $I_b=I(R_b)$ by:
\begin{equation}
\label{Iprime}
I^{\prime}=I_b 2^{-\gamma/\alpha} \exp \left[ b_n \left( 2^{1/\alpha} r_b/r_e \right)^{1/n} \right],
\end{equation}
and $b_n \approx 1.992n-0.3271$ (e.g., \citealt{Gra2005}). The core-S\'{e}rsic fits for five representative galaxies from our sample, arranged top to bottom from brightest to faintest, are illustrated in Figure~\ref{fig:profiles}\emph{a-e} by the solid black lines; the S\'{e}rsic component of the fits are highlighted by the dot-dashed blue lines. The progression from central light deficit to excess shown is characteristic of our sample, although it should be noted that there are also a small number ($\lesssim 10\%$ of our sample) of fainter galaxies ($M_B \gtrsim -17.5$~mag) which do not deviate significantly from a single S\'{e}rsic model at small radii. These ``non-nucleated dwarf" galaxies are discussed in \S4 of \citetalias{Cot2007}.

\subsection{Deprojecting the Surface Brightness Profiles}

Under the assumption of spherical symmetry, the surface brightness profile $I(R)$ of a galaxy can be deprojected into the luminosity density profile $j(r)$ using an Abel integral,
\begin{equation}
\label{eq: j(r)}
j(r) = -\frac{1}{\pi} \int_r^{\infty} \frac{dI}{dR} \frac{dR}{\sqrt{R^2 -r^2}},
\end{equation}
where $r$ denotes the radius of the galaxy in spherical coordinates and $R$ denotes the radius of the galaxy projected onto the sky (e.g., \citealt{Bin1987}). 

For each program galaxy (excepting five for which we were unable to perform a core-S\'{e}rsic fit due to the presence of dust) the above integration was performed numerically using $I(R)$ as specified in \S \ref{sec: Parameterization} and using surface brightness fluctuation distances from \citet{Bla2009}. The deprojections of five representative galaxies are shown by the solid black lines in the rightmost panels in Figure~\ref{fig:profiles}. The deprojections were carried out a second time excluding the central power law (i.e., using the S\'{e}rsic component of the surface brightness profile fit only). This is plotted as the blue curves in panels ($k$-$o$) of Figure~\ref{fig:profiles}. In order to verify our deprojection routine, each luminosity density profile was ``re-projected'' numerically to confirm agreement with the original (parameterized) surface brightness profile. These re-projections were found to match the original profiles very closely,  as shown by the green dashed lines in Figure~\ref{fig:profiles}\emph{f-j}.

\section{Results}
\label{sec: Results}

\begin{figure}
\includegraphics[width=\textwidth]{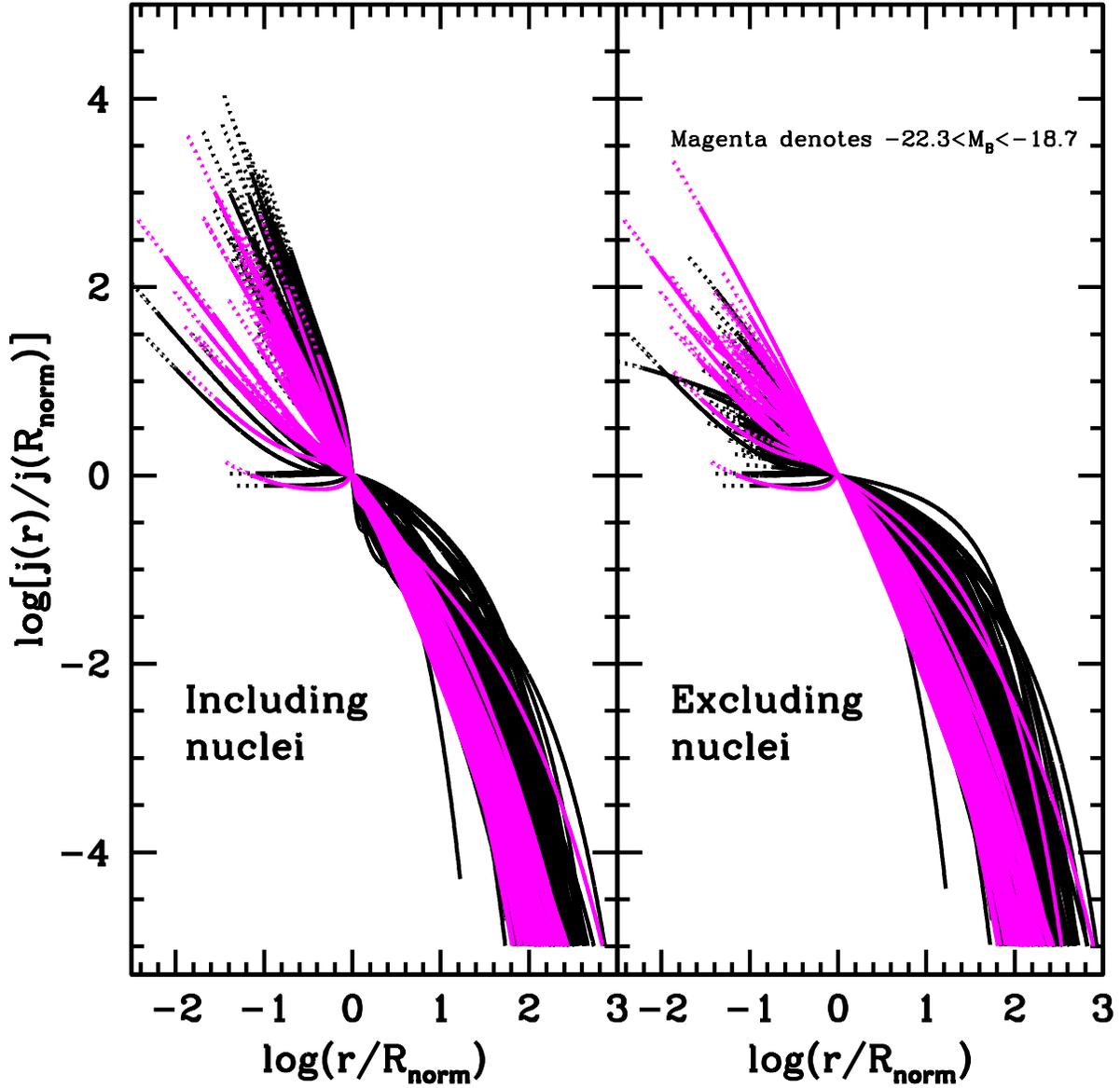}
\caption{Luminosity density profiles of the ACSVCS and ACSFCS galaxies in the $z$-band, scaled to the break radius $R_b$ and the luminosity density at this radius, both including nuclei (\emph{left panel}) and excluding nuclei (\emph{right panel}). Each profile is plotted as a solid line down to the approximate resolution limit ($0\Sec1$) and as a dotted line down to the pixel size for the ACS/WFC ($0\Sec05$). Galaxies ranging from $22.3$ mag $\la M_B \la 18.7$ mag are drawn in magenta. The remainder of the galaxies are shown in black.}
\label{fig:norm}
\end{figure}

The luminosity density profiles of the ACSVCS and ACSFCS galaxies are plotted in Figure~\ref{fig:norm}, normalized as a function of $r/R_b$ in order to compare all $138$ galaxies at once. The panel on the left shows the deprojections of the full core-S\'{e}rsic profiles while the one on the right shows the deprojections excluding nuclei. (In other words, for those galaxies with central light \emph{excesses}, only the deprojected S\'{e}rsic component of the outer surface brightness profile fits are shown on the right whereas galaxies with central light \emph{deficits} are deprojected from the core-S\'{e}rsic fits including the central power-law component in \emph{both} panels.)  Galaxies brighter than $M_B=-18.7$~mag have been highlighted in magenta: these galaxies span  a 3.6~mag range centered at $M_B=-20.5$~mag, which marks the reported separation between the core and power-law galaxies (e.g., \citetalias{Lau2007}). For the sake of simplicity, we show the profiles obtained from the $z$-band observations only, given that the overall results using the $g$-band observations are the same. 

The left panel of Figure~\ref{fig:norm} is analogous to Figure~3 of \citetalias{Geb1996} and Figure~6 of \citetalias{Lau2007}, who numerically deprojected non-parametric representations (which therefore include nuclei) of the surface brightness profiles of a heterogeneous sample of galaxies drawn from the \emph{HST} archive. 
Whereas they find a clear region in which the inner profiles do not fall, the inner profiles in Figure~\ref{fig:norm} fan out to cover a continuous range of slopes. This is true whether we look at the entire sample or only at those galaxies with magnitudes around $M_B\sim-20.5$ mag. It is also true whether nuclei are included or not. Note that a small gap in the inner slopes of the luminosity density profiles in the left panel may be perceived. We do not, however, believe this is significant, for several reasons. Firstly, the gap is so small that one single galaxy with the appropriate central slope would fill it in. Secondly, this gap is nowhere near the magnitude of the gaps seen by \citetalias{Geb1996} and \citetalias{Lau2007}. Lastly, and most importantly, as we show in what follows, no quantitative analysis of the inner slopes of these profiles reveals a discontinuity with galaxy magnitude.

To this end, in order to further characterize the behavior of the inner luminosity density profiles, we compute the logarithmic derivative $\gamma_{\rm 3D}$ of $j(r)$ numerically at different fractions (0.5\%, 1\%, 5\%, and 30\%) of the effective radius $R_e$, where $\gamma_{\rm 3D}(r)=-\frac{d \log j(r)}{d \log r}$. The resulting values of $\gamma_{\rm 3D}$ are given for the Virgo galaxies in Table~\ref{tab:acsvcs} and for the Fornax galaxies in Table~\ref{tab:acsfcs}.

\begin{figure}
\includegraphics[width=\textwidth]{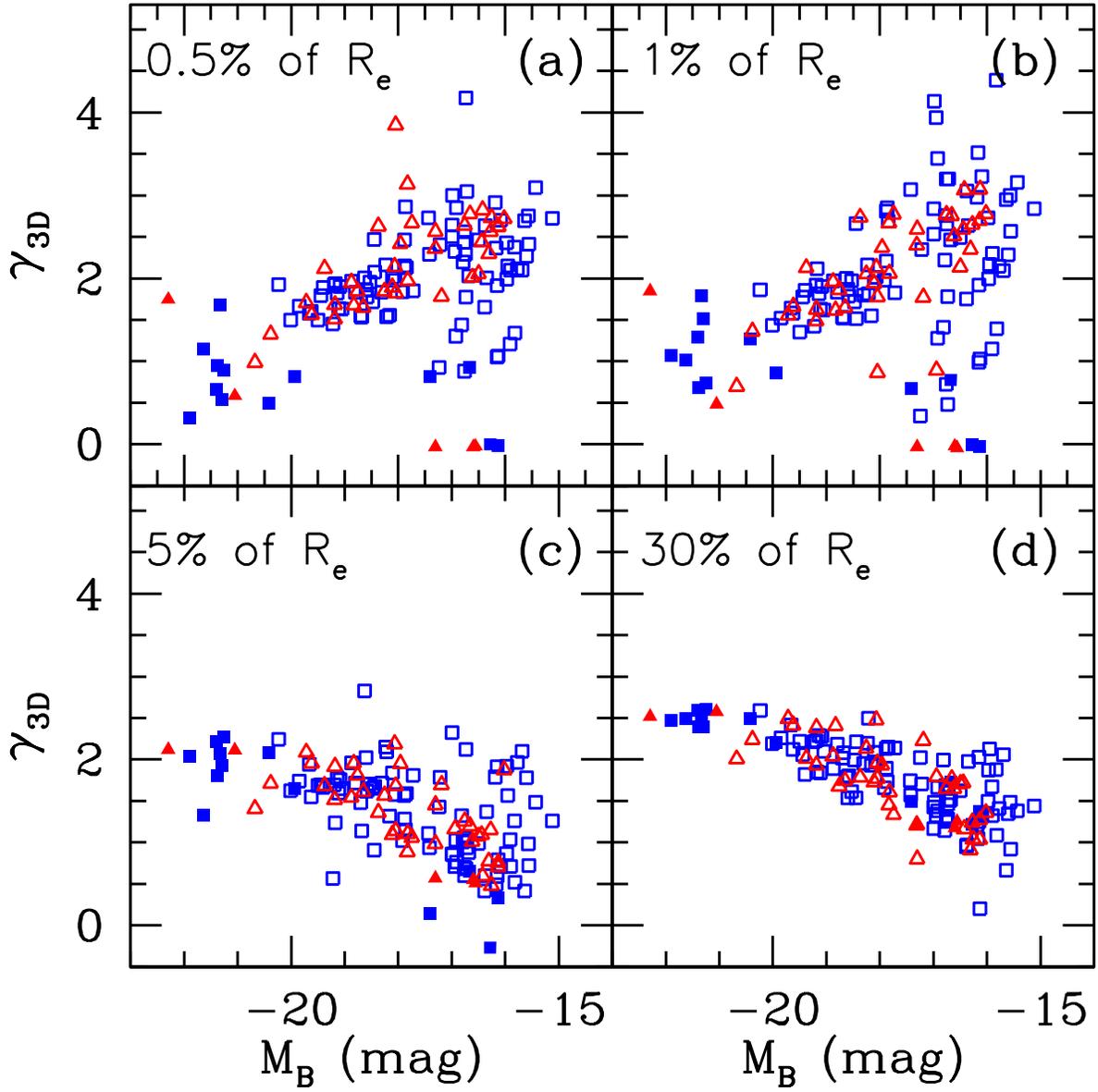}
\caption{Logarithmic slopes $\gamma_{\rm 3D}$ of the $z$-band luminosity density profiles --- with nuclei included --- as a function of $M_B$ at 0.5\%, 1\%, 5\%, and 30\% of the effective radius $R_e$ for the Virgo (\emph{blue squares}) and Fornax (\emph{red triangles}) galaxies in our sample. The open points indicate nucleated galaxies. Note the systematic steepening of the central brightness profiles (i.e., at $0.5\%$ and $1\%$ of $R_e$) due to the presence of nuclei and smooth transition with decreasing galaxy luminosity.}
\label{fig:gammaMB}
\end{figure}

\begin{figure}
\includegraphics[width=\textwidth]{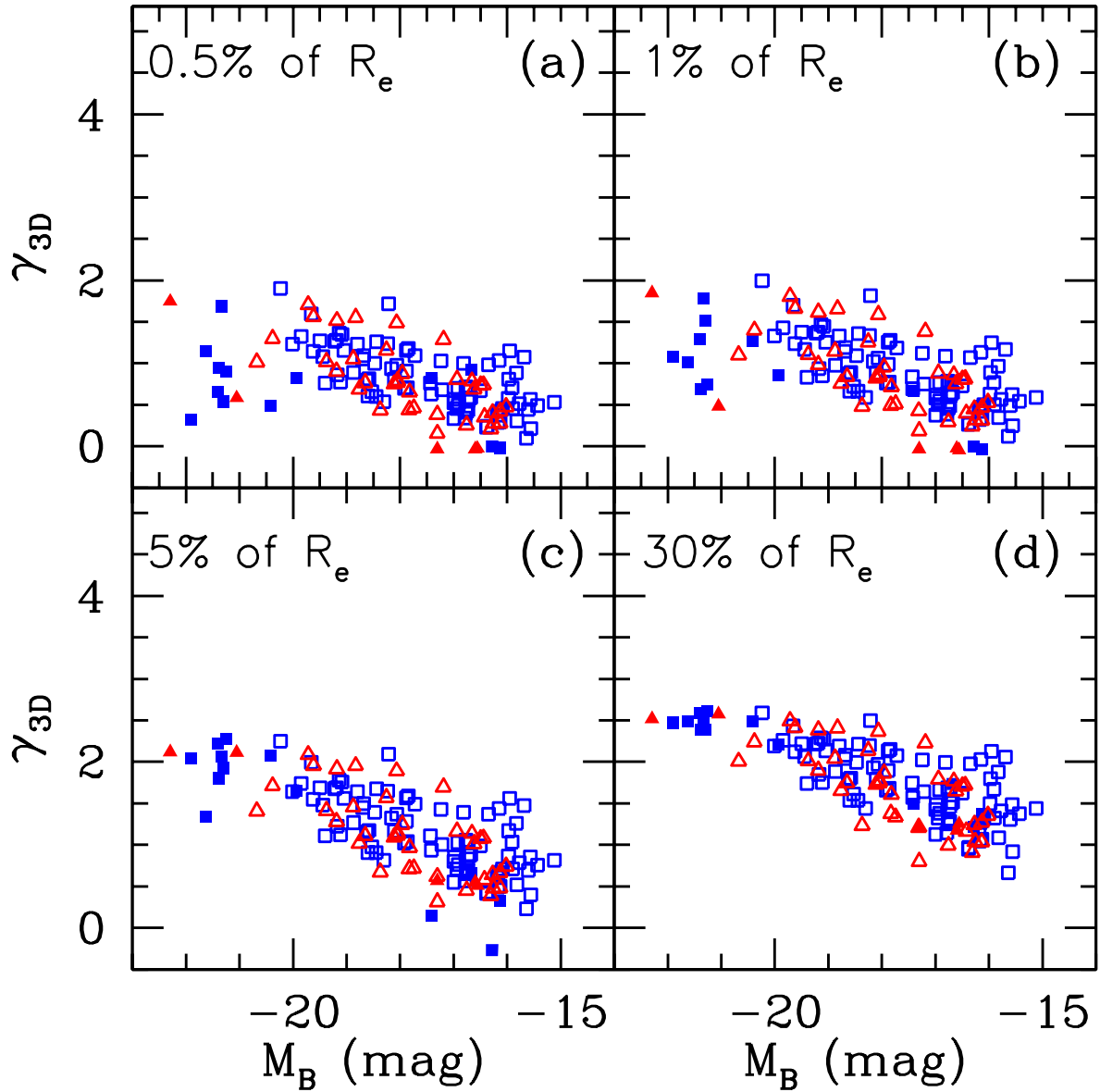}
\caption{Same as Figure~\ref{fig:gammaMB} except that nuclei, when present, are excluded.}
\label{fig:gammaMBnonuc}
\end{figure}

Figures~\ref{fig:gammaMB} and \ref{fig:gammaMBnonuc} plot $\gamma_{\rm 3D}$  as a function of $M_B$ for the $z$-band, including and excluding nuclei, respectively. 
As expected, the nuclei do not impact the distributions when the slope is measured beyond 5\% of $R_e$: most stellar nuclei have effective radii of $\sim2\%$ of $R_e$ and their contribution at much larger radii is therefore negligible.  There is no evidence of {\it disjoint} populations in these figures, although the distributions have large scatter (especially when nuclei are included and the slope is measured at small radii, since under these conditions the gradients in the profiles can change dramatically), are not linear, and show clearly varying systematic trends as one moves down the luminosity function. For instance, in the brightest galaxies the central brightness profile (projected or not) flattens as galaxies become brighter, while the opposite is true for fainter systems if the nuclei are excluded. (Indeed, as noted by \citealt{Fer2006}, once we subtract the nuclei, the shallowest inner profiles belong to the \emph{faintest} systems.) Likewise, while the fainter galaxies define nearly linear relations in all plots, the brighter galaxies deviate from the extrapolation of such relations. However, non-linearity should not be mistaken as evidence of separate populations. As one moves from the brightest to the faintest systems, the slopes change smoothly and continuously: in none of the panels do the points segregate into separate regions of the parameter space.  

\begin{figure}
\includegraphics[width=\textwidth]{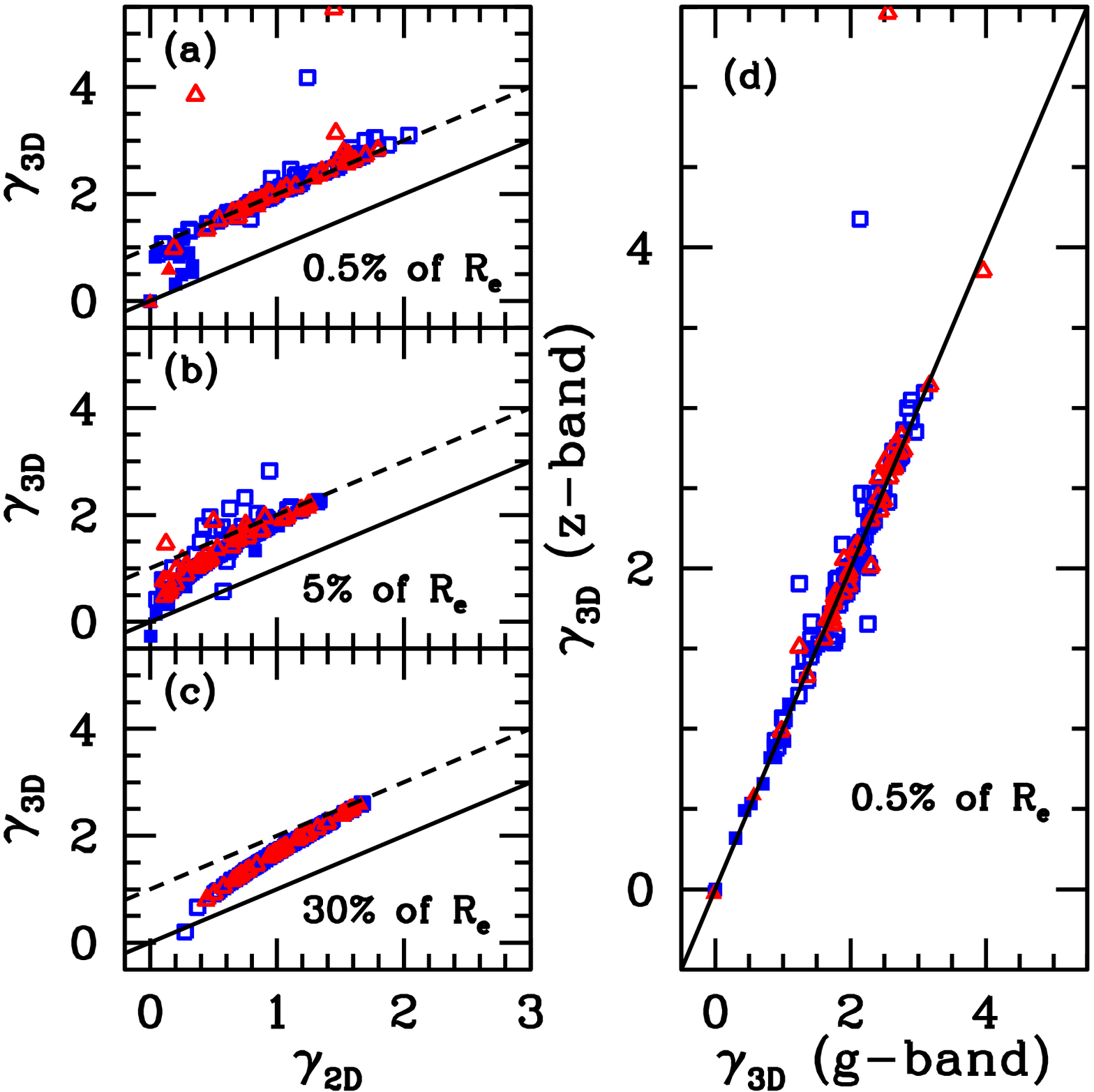}
\caption{(\emph{Panels~a-c}) Comparison of the logarithmic slopes of the $z$-band luminosity density profiles $\gamma_{\rm 3D}$ to those of the surface brightness profiles $\gamma_{\rm 2D}$. The comparison is shown for slope measurements made at 0.5\%,  5\%, and 30\% of the effective radius.  ACSVCS galaxies are denoted by blue squares and ACSFCS galaxies by red triangles. Galaxies with nuclei are indicated by the open symbols. The solid lines are for $\gamma_{\rm 3D}=\gamma_{\rm 2D}$ and the dashed lines are for $\gamma_{\rm 3D}=\gamma_{\rm 2D}+1$.
(\emph{Panel~d}) Comparison of the luminosity density slopes measured at 0.5\% of $R_e$ for the $g$- and $z$-band profiles. The solid line shows a one-to-one relation.}
\label{fig:gzcompare}
\end{figure}

A comparison of $\gamma_{\rm 3D}$ and its two-dimensional analog, $\gamma'=\gamma_{\rm 2D}=-d\log I/d\log R$ (in the $z$-band), as well as a comparison of $\gamma_{\rm 3D}$ in the $g$ and $z$-bands, is shown in Figure~\ref{fig:gzcompare}\emph{a-d}. (The nuclei have been included in this figure.)  As expected (e.g., \citealt{Deh1993, Mer1996}; \citetalias{Geb1996}),  $\gamma_{\rm 3D} \gtrsim1$ correspond to projected slopes $\gamma_{\rm 2D}$ that are shallower by $\sim1$ dex, whereas galaxies spanning the range of slopes $0 \la \gamma_{\rm 3D} \la 1$ all translate, once projected, to surface brightness profiles with nearly identical shallow cores characterized by $\gamma_{\rm 2D}\sim0$, a point to which we will return later. Panel (\emph{d}) of Figure~\ref{fig:gzcompare}, which compares the results obtained using the different bandpasses at 0.5\% of $R_e$, demonstrates that there is good agreement between the $g$ and $z$ profiles. Taken together with Figures~\ref{fig:gammaMB} and \ref{fig:gammaMBnonuc}, there is no evidence of a separation of galaxies into distinct classes claimed by \citetalias{Geb1996} and \citetalias{Lau2007}. 

\begin{figure}
\includegraphics[width=\textwidth]{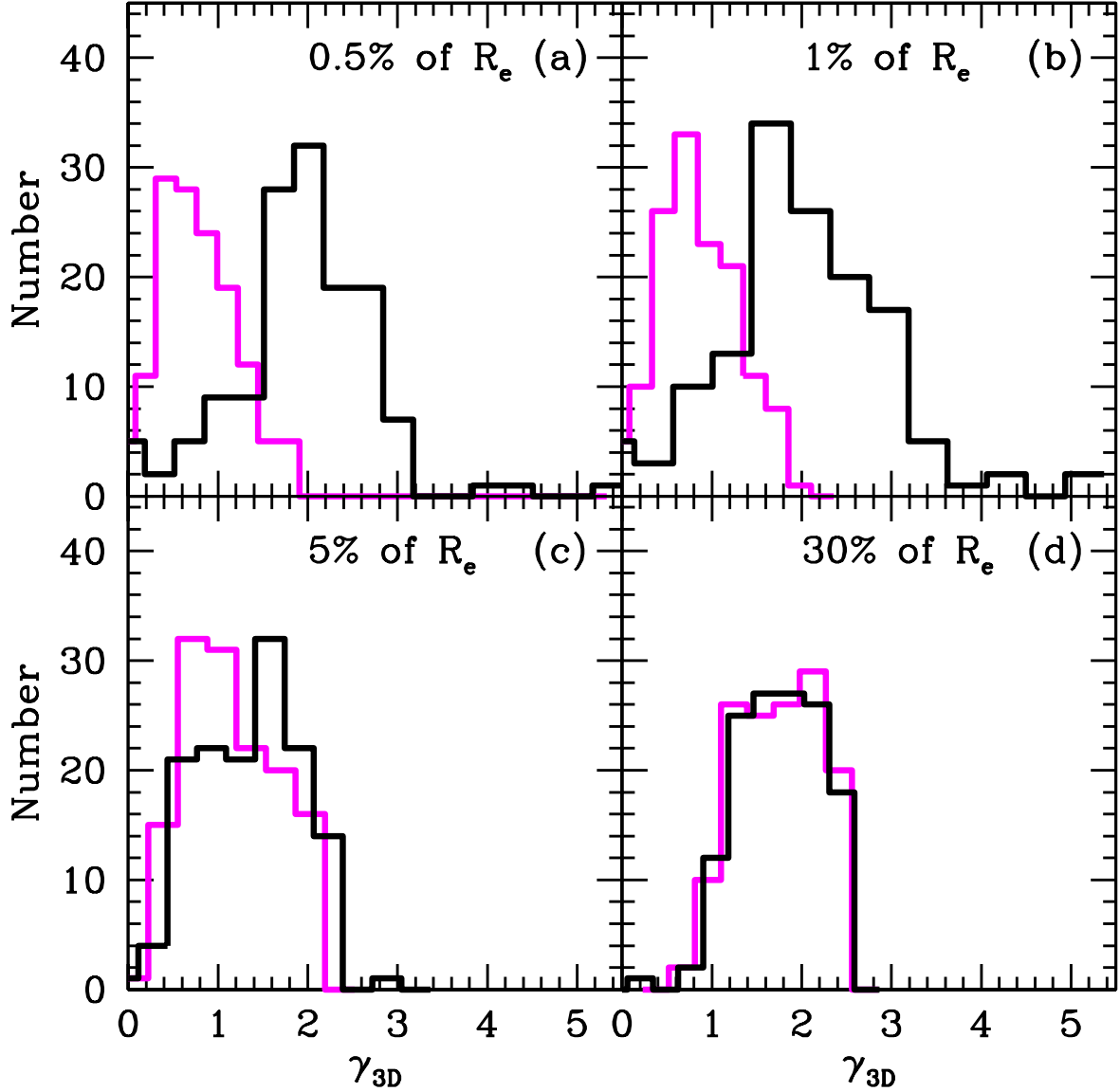}
\caption{Histograms showing the distribution of $\gamma_{\rm 3D}$ at $0.5\%$, $1\%$, $5\%$, and $30\%$ of the effective radius for each galaxy. This figure illustrates how, generally, the distribution changes depending on where $\gamma_{\rm 3D}$ is measured. The black histograms denote values for $\gamma_{\rm 3D}$ including nuclei (where they are present) while the magenta ones show the distribution of $\gamma_{\rm 3D}$ without nuclei. Note that the largest values of  $\gamma_{\rm 3D}$ seen for the galaxies in the black histogram in panels \emph{a-c} disappear when nuclei are excluded because these measurements reflect the logarithmic slopes of the nuclei, not the underlying galaxy.}
\label{fig:gammahist}
\end{figure}

\begin{figure}
\includegraphics[width=\textwidth]{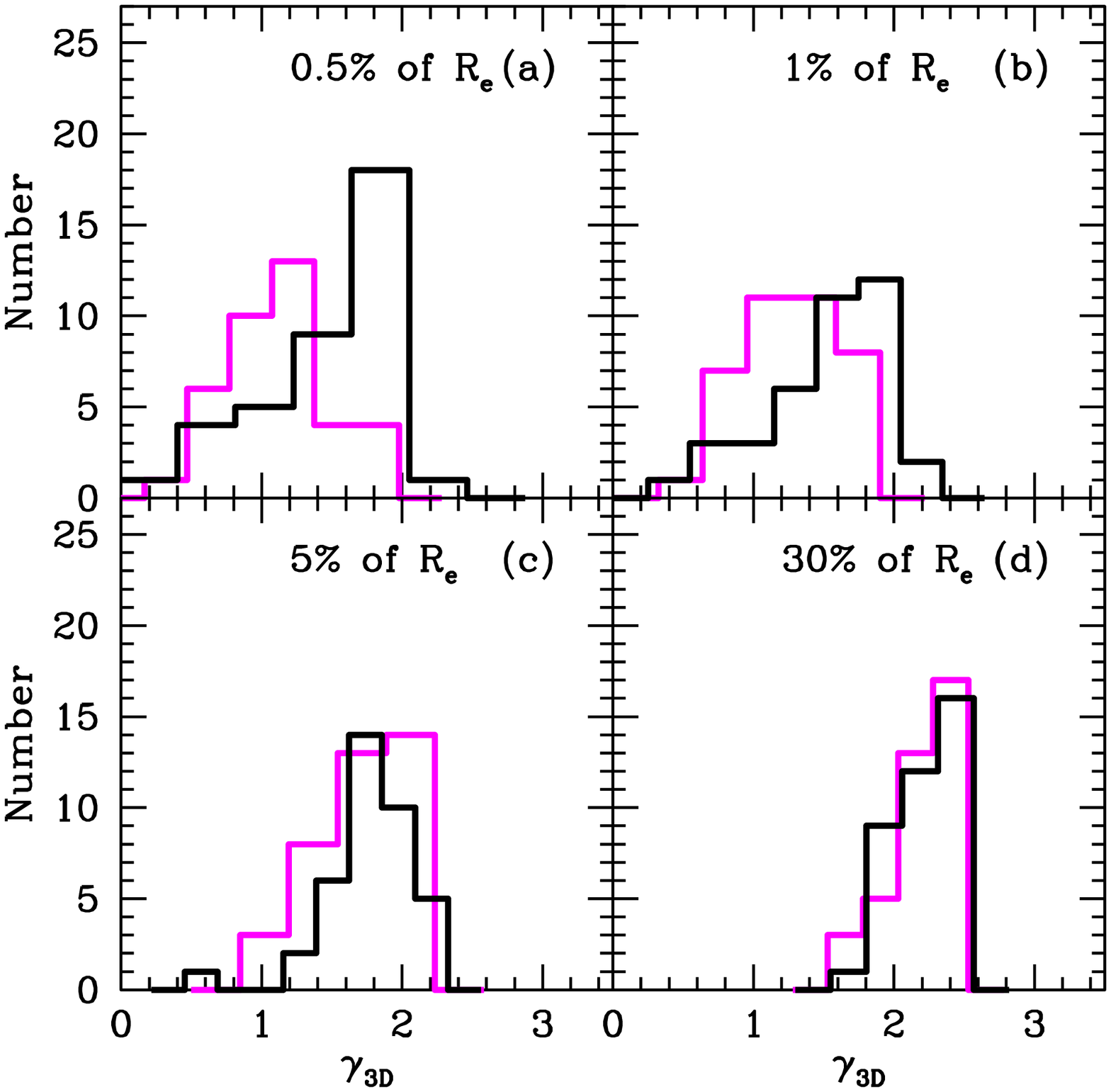}
\caption{Same as Figure~\ref{fig:gammahist}, except only galaxies within $\sim \pm2$~mag of $-20.5$~mag (i.e., the magnitude which has been claimed to separate ``core'' and ``power-law'' galaxies) are included.}
\label{fig:gammahist_minus20.5}
\end{figure}

Finally, histograms of the $\gamma_{\rm 3D}$ distributions at various fractions of $R_e$ are shown in Figure~\ref{fig:gammahist}\emph{a-d}, both including (\emph{black}) and excluding (\emph{magenta}) compact stellar nuclei, as well as in Figure~\ref{fig:gammahist_minus20.5}\emph{a-d} for galaxies with $-22.3$~mag~$\la M_B \la -18.7$~mag (a range, as mentioned earlier, centered on $M_B\sim-20.5$~mag, where the dichotomy is supposed to occur). The ACSVCS and ACSFCS samples are combined in both figures as the histograms of each sample individually do not differ in substance. The bin size is the  ``optimal'' bin size  given by \citet{Ize1991} as $2 ({\rm IQR}) N^{-1/3}$, where $N$ is the total number of objects and  $\rm{IQR}$ is the interquartile range, i.e., the range of the second and third quartiles of the ordered $\gamma_{\rm 3D}$ distribution. These figures confirm the impression, drawn from Figures~\ref{fig:gammaMB} and \ref{fig:gammaMBnonuc}, that the distributions of $\gamma_{\rm 3D}$ are not bimodal. Note also that the distribution of $\gamma_{\rm 3D}$ is dependent on where $\gamma_{\rm 3D}$ is measured, as expected given that, in S\'{e}rsic models, the curvature itself is a function of radius.

By comparing the histograms showing the slopes obtained including and excluding the nuclei in Figure~\ref{fig:gammahist}, it is evident that the large values of $\gamma_{\rm 3D}$ observed at the innermost radii reflect the presence of nuclei: not only do the largest values disappear in the histogram excluding the nuclei, but also, the histograms including and excluding the nuclei become very similar when the slopes are measured at larger radii, past the point at $\sim2\%R_e$ where the underlying galaxy begins to dominate the profile. All in all, and most importantly, the distributions do not appear bimodal, regardless of whether or not the nuclei are included and whether the entire sample, or just a limited magnitude range around the point where a minimum is supposed to occur, are plotted.

\section{Caveats and Comparisons with Previous Work}
\label{sec: Caveat}

The results presented above hinge on two main assumptions: (1) that the adopted parameterization --- which, once convolved with the instrumental PSF, describes the observed surface brightness profile quite well --- is a good description of the intrinsic (pre-convolution) projected profile; and (2) that the galaxies can be deprojected under the assumption of sphericity. In their analyses, \citetalias{Geb1996} and \citetalias{Lau2007} performed numerical deprojections of  non-parametric spline fits to the surface brightness profiles derived by fitting isophotes to deconvolved images; in this paper, we have opted to perform a numerical deprojection of the parametric models that, once convolved with the instrumental PSF, best fit the surface brightness profiles derived by fitting isophotes to PSF-convolved images. 

Both methods have disadvantages. Deconvolution of noisy data does not generate images with infinitely high spatial resolution, and therefore, pre-convolution images are not generally coincident with deconvolved images. Noise amplification in deconvolved images can also be a concern. Additionally, using a parametric approach, the nucleus can be either included or excluded, whereas this cannot be easily accomplished using the deprojection method of \citetalias{Geb1996} and \citetalias{Lau2007}. (Indeed, nuclei are included in their analysis.) Conversely, because the nuclear component is often just barely resolved in the images, a given parameterization may not be unique. A priori, therefore, there is no obvious reason to prefer one approach to the other. 

\begin{figure}
\begin{center}
\includegraphics[width=6 in]{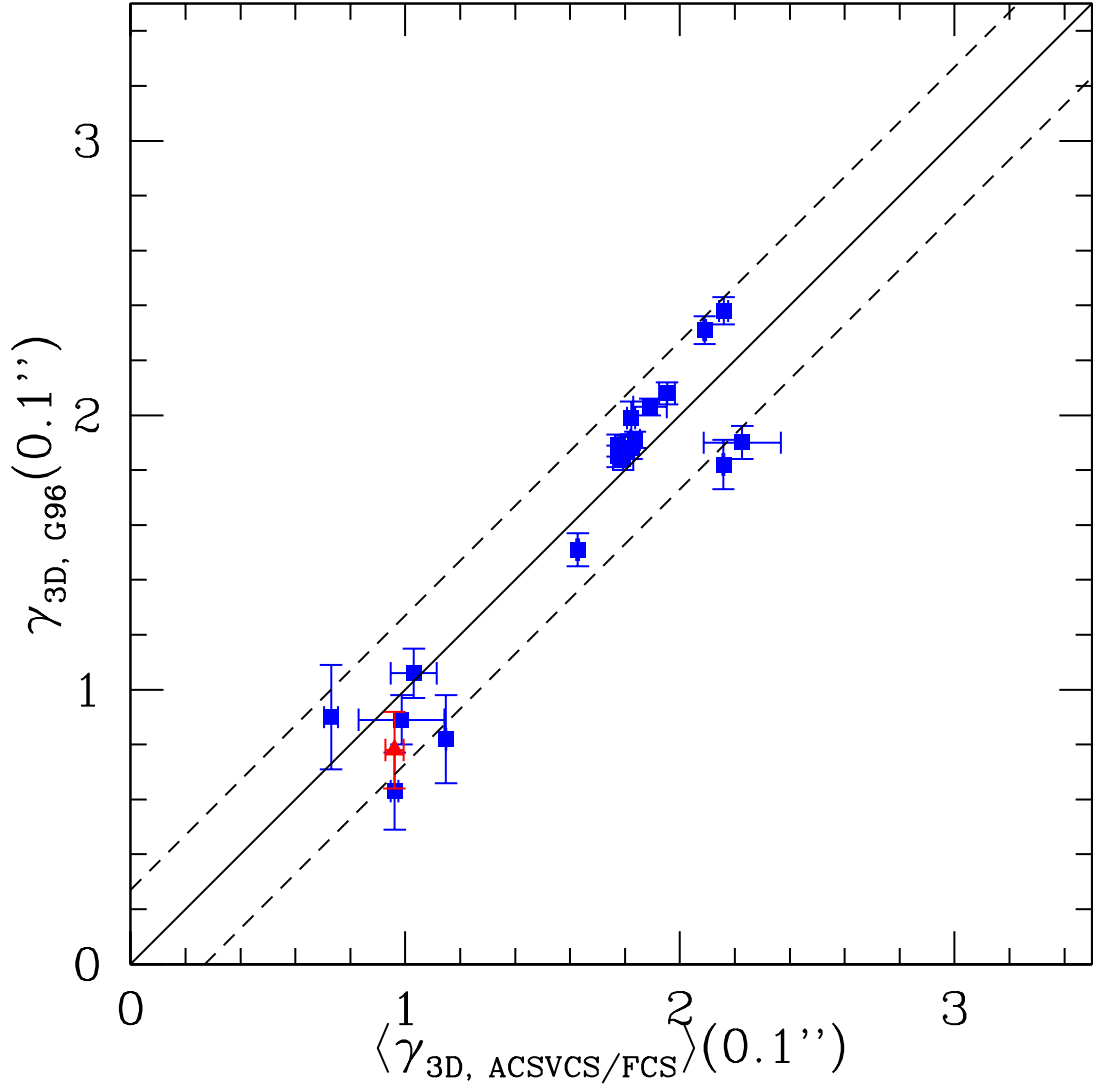}
\caption{A comparison of our values for $\gamma_{\rm 3D}$ at $0\Sec1$ (averaged between the $g$ and $z$-band profiles derived from the ACS images) with those in common with \citetalias{Geb1996} (who used WFPC1 data). The blue squares are ACSVCS galaxies and the red triangle is FCC~213, the only ACSFCS galaxy included in \citetalias{Geb1996}'s analysis. The error bars are as stated in \citetalias{Geb1996} for the vertical axis and represent the spread in the $g$ and $z$-band measurements of $\gamma_{\rm 3D}$ for the horizontal axis. The solid black line denotes where the two values are equal and the dashed lines represent a $2\sigma$ deviation from this line.}
\label{fig:gammaGeb} 
\end{center}
\end{figure}

We have compared our $\gamma_{\rm 3D}$ values for the 19 galaxies (14 classified as E, 3 as E/S0, and 2 as S0) in common between our sample and that of \citetalias{Geb1996}, spanning a magnitude range between $-22 \lesssim M_B \lesssim -15.7$~mag (although most are brighter than $M_B\sim-18.5$~mag). This is shown in Figure~\ref{fig:gammaGeb}. For this comparison, our measurements are made at $0\Sec1$, the fixed angular radius adopted for the slope measurements presented in \citetalias{Geb1996}. The values we compute for $\gamma_{\rm 3D}(0\Sec1)$ are given in Tables~\ref{tab:acsvcs} and \ref{tab:acsfcs} for our entire sample. (We are unable to compare with \citetalias{Lau2007} because they did not tabulate their values for $\gamma_{\rm 3D}$.) The systematic difference between our slopes and those of \citetalias{Geb1996} amounts to an average of only $\sim5\%$ of $\gamma_{\rm 3D}$, with no obvious trend with galaxy magnitude. We therefore conclude that the choice of image processing (whether or not to deconvolve the images) and fitting functions (core-S\'{e}rsic versus spline) has little effect on the derived slopes and is not sufficient to alter any of our conclusions.

The consequences of our assumption of sphericity are more difficult to assess. As detailed in \citet{Fer2006}, the observed one-dimensional surface brightness profiles used to carry out the model fitting are given as a function of the ``geometric mean radius", $R = a[1-\epsilon(a)]^{1/2}$, where $a$ is the radius measured along the isophotal semimajor axis and $\epsilon$ is the ellipticity at $a$. Using the geometric mean radius instead of the semi-major axis has the effect of ``circularizing" the isophotes on the plane of the sky. In this sense, the deprojected profiles represent a characteristic luminosity density along an axis that is intermediate between the observed (projected) major and minor axes of the galaxy. 

The assumption that the galaxy is circularly symmetric perpendicular to the plane of the sky is born out of necessity, as the inclination angle of the galaxy is generally unknown. The expectation, however, is that when looking at a sample of randomly oriented galaxies, our ``edge-on" approximation would lead us to overestimate the luminosity density and inner slopes (by amounts which depend on the inclination and on the shape of the observed profiles). Qualitatively, this is unlikely to reconcile our results with the existence of a core/power-law dichotomy because luminous (``core'') galaxies are traditionally believed to be close to spherically-symmetric, pressure-supported systems, while fainter (``power-law'') galaxies are believed to be flattened (e.g., \citealt{Lau2007}; but see also \citealt{Ems2007}). Relaxing our edge-on assumption would therefore have little affect on the core galaxies, but would {\it decrease} the luminosity density slope of power-law galaxies, therefore narrowing, rather than widening, any gap between the two. In any case, both our analysis and that of \citetalias{Geb1996} and \citetalias{Lau2007} assume sphericity. This assumption, therefore, could not account for the different conclusions reached by their studies and ours.

\begin{figure}
\begin{center}
\includegraphics[width=6 in]{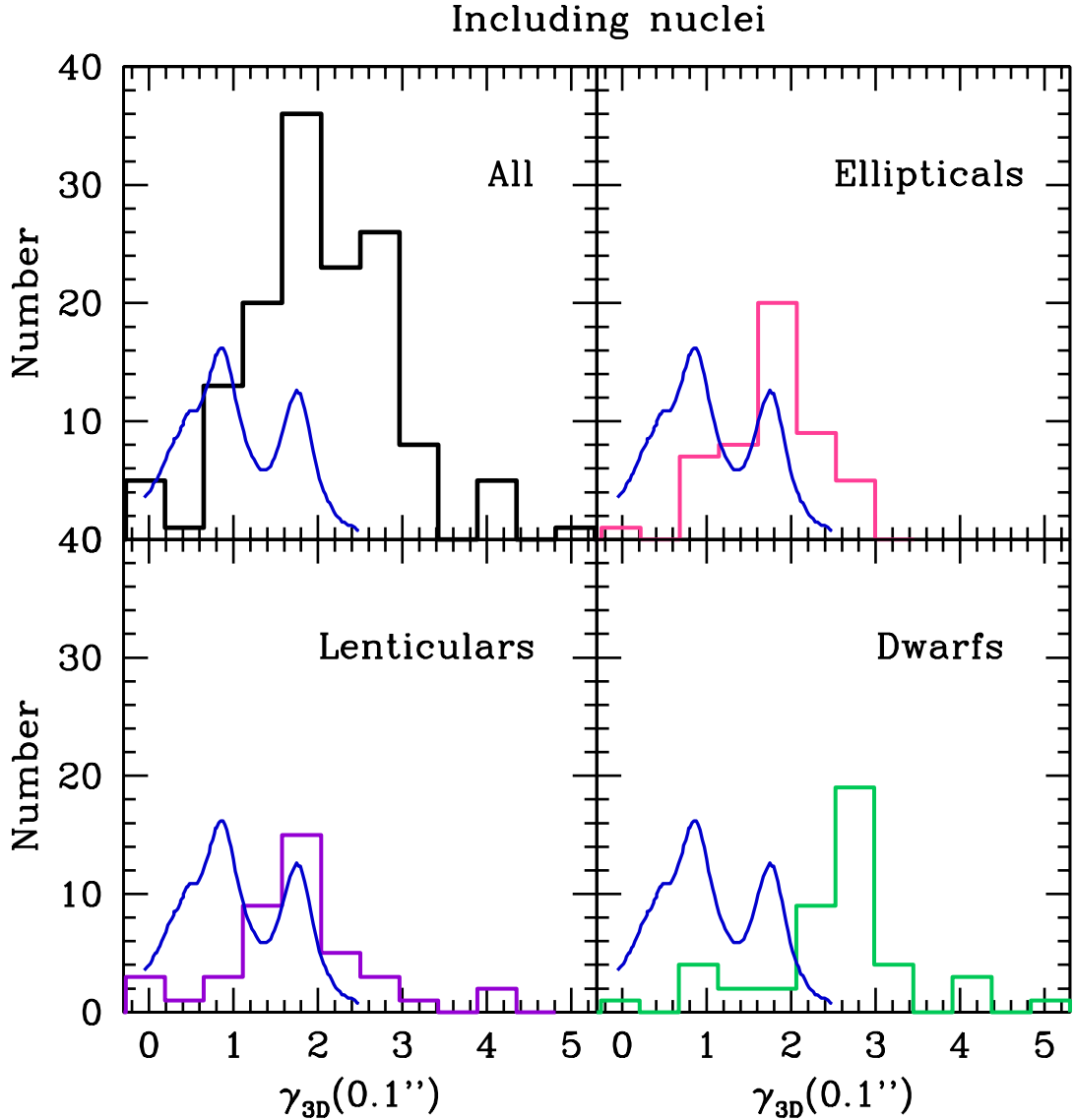}
\caption{Histograms of the values computed for $\gamma_{\rm 3D}$ at $0\Sec1$, including nuclei, for comparison with \citetalias{Geb1996} and \citetalias{Lau2007} (as it is at this radius -- the approximate resolution of \emph{HST} -- at which they both measure $\gamma_{\rm 3D}$). The black histogram includes all galaxies, while the colored histograms separate the galaxies by morphological type (from P.~C\^ot\'e et al. 2011, \emph{in prep}): pink denotes ellipticals; purple, lenticulars; and green, dwarfs. The blue distribution is the fit to \citetalias{Lau2007}'s $\gamma_{\rm 3D}(0\Sec1)$ distribution, as shown in their Figure 7, second panel down. The bimodality they observe is not present in our sample.}
\label{fig:histpoint1} 
\end{center}
\end{figure}

\begin{figure}
\begin{center}
\includegraphics[width=6 in]{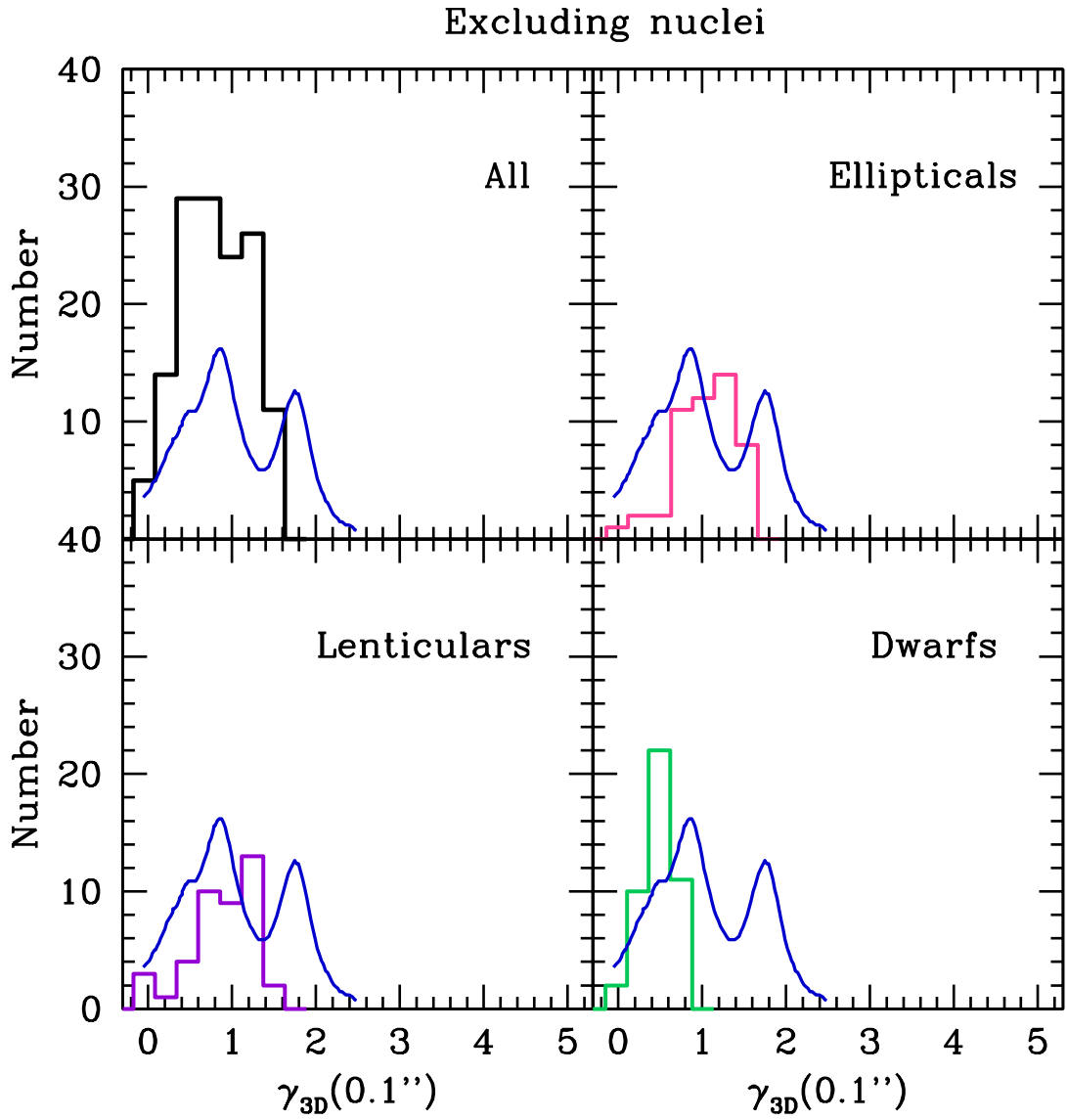}
\caption{Same as Figure~\ref{fig:histpoint1} except that nuclei, when present, are excluded. Note that nuclei are included in the distribution of \citetalias{Lau2007}, shown by the blue curve.}
\label{fig:histpoint1_nonuc} 
\end{center}
\end{figure}

\begin{figure}
\includegraphics[width=\textwidth]{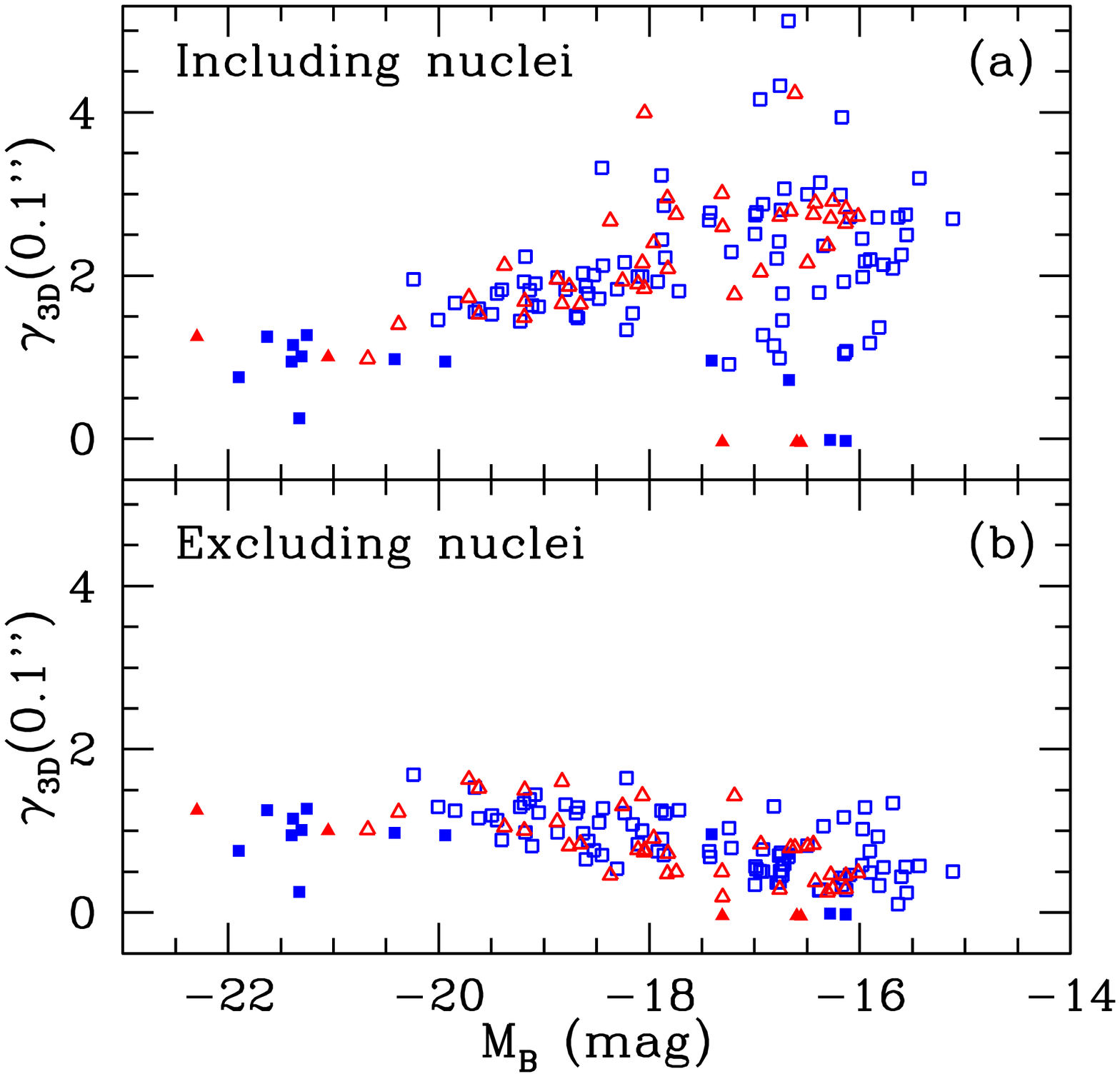}
\caption{Logarithmic slopes $\gamma_{\rm 3D}(0\Sec1)$ as a function of $M_B$ from the $z$-band luminosity density profiles for the Virgo (\emph{blue squares}) and Fornax (\emph{red triangles}) galaxies in our sample. The open points indicate nucleated galaxies. There is a smooth transition in $\gamma_{\rm 3D}$ with galaxy luminosity, regardless of whether one considers (\emph{panel a}) the slopes including nuclei or (\emph{panel b}) the slopes excluding nuclei.}
\label{fig:gammaMB_point1as}
\end{figure}

The disagreement between our results and those of \citetalias{Lau2007}, specifically, is made clear by the histograms in Figures~\ref{fig:histpoint1} and \ref{fig:histpoint1_nonuc}, which plot the distribution of $\gamma_{\rm 3D}(0\Sec1)$ both including and excluding nuclei. Figure~\ref{fig:histpoint1}, which plots $\gamma_{\rm 3D}(0\Sec1)$ including nuclei, is directly analogous to the second panel of \citetalias{Lau2007}'s Figure~7 as they include nuclei in their analysis. For completeness, we also show our $\gamma_{\rm 3D}(0\Sec1)$ distributions excluding nuclei in Figure~\ref{fig:histpoint1_nonuc}. In the past, there has been some debate as to which morphological types should be included in this type of analysis. In a companion paper (P.~C\^ot\'e et al. 2011, \emph{in prep}), we will present a detailed comparison of morphological classifications for ACSVCS and ACSFCS galaxies from a variety of sources, including \citet{Bin1985}, the RC3 \citep{deV1991}, \citet{Fer1989}, and new classifications by S.~van~den~Bergh made directly from the ACS imaging used in this paper. Although there is, at best, only fair agreement among the different studies, we have used these classifications to assign ``consensus morphologies'' (E, S0, or dwarf) to the complete sample of galaxies. While these morphologies are often quite uncertain for individual galaxies, classifying them into three broad bins for ellipticals, lenticulars, and dwarfs (which we take to include both dEs and dS0s) allows an examination of how our results might change with different morphological samples. At the same time, we emphasize that the distinction between, e.g., Es and S0s is often highly ambiguous as these galaxies follow the same global scaling relations (e.g., \citealt{Fer2006}; P.~C\^ot\'e et al. 2011, \emph{in prep}) and can often not be differentiated even with the addition of kinematic information (e.g., \citealt{Ems2007, Chi2009}).\footnote{In fact, it is for these reasons, as well as because disk fitting functions are degenerate -- meaning that, although exponential profiles are generally assumed, there is no \emph{a priori} reason to prefer this functional form to another -- that we feel it is justified not to perform a bulge-disk decomposition in our analysis. In any case, neither \citetalias{Geb1996} nor \citetalias{Lau2007} perform a decomposition for the lenticular galaxies in their samples, so this does not explain our differing results.} Nevertheless, Figures~\ref{fig:histpoint1} and \ref{fig:histpoint1_nonuc} show the results of dividing the sample in this way into ellipticals (in pink), lenticulars (in purple), and dwarfs (in green).  Whereas \citetalias{Lau2007} find peaks in the distribution at $\gamma_{\rm 3D}\thickapprox0.9$ and $\gamma_{\rm 3D}\thickapprox1.8$, and a trough at $\gamma_{\rm 3D}\thickapprox1.4$ -- as shown by the fit to \citetalias{Lau2007}'s $\gamma_{\rm 3D}(0\Sec1)$ distribution, plotted over our distributions in Figures~\ref{fig:histpoint1} and \ref{fig:histpoint1_nonuc} -- none of our distributions support this type of dichotomy. For completeness, we have also plotted $\gamma_{\rm 3D}(0\Sec1)$ versus $M_B$ in Figure~\ref{fig:gammaMB_point1as}, both including and excluding nuclei. It clearly shows a \emph{gradual trend}  in $\gamma_{\rm 3D}(0\Sec1)$ with magnitude, not a discontinuous steepening. 

\begin{figure}
\includegraphics[width=\textwidth]{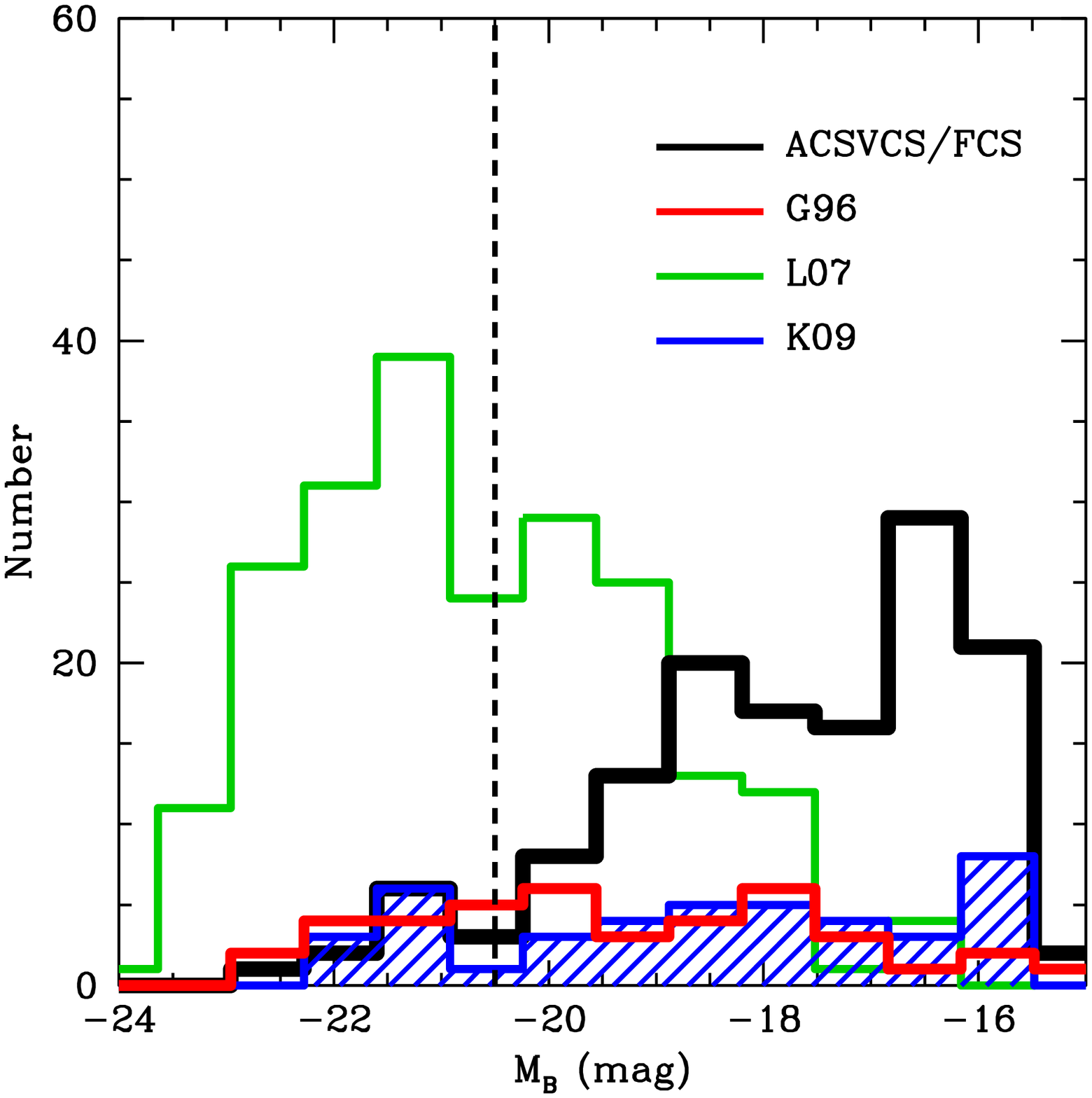}
\caption{Luminosity functions of the ACSVCS/FCS (i.e., the sample used in this paper; \emph{black}), the \citetalias{Geb1996} sample (\emph{red}), the \citetalias{Lau2007} sample (\emph{green}), and the \citetalias{Kor2009} sample (\emph{blue with hatching}). Note that the ACSVCS/FCS is complete down to $M_B\sim-19.2$, more than a magnitude fainter than  $M_B\sim-20.5$~mag, where the divide between core and power-law galaxies is supposed to occur (indicated by the dashed line).}
\label{fig:lumfunc}
\end{figure}

The origin of the discrepancy with the results of both \citetalias{Geb1996} and \citetalias{Lau2007} lies in the samples analyzed in the different studies. As \citetalias{Cot2007} explain in more detail, our sample consists of early-type galaxies in Virgo and Fornax that have a well-understood selection function, one that closely resembles the 
Schechter-type luminosity function exhibited by galaxies in general and in these clusters specifically. By contrast, the samples analyzed by \citetalias{Geb1996} and \citetalias{Lau2007} are strongly overabundant in bright (``core'') galaxies compared to standard luminosity functions and quickly become incomplete at fainter magnitudes -- a luminosity regime where the ACSVCS and ACSFCS samples show clearly that the class of early-type so-called ``power-law'' galaxies nearly always have a two-component structure: i.e., the galaxy itself and a nuclear component. Figure~\ref{fig:lumfunc} plots the luminosity functions for the combined ACSVCS/FCS,  \citetalias{Geb1996} , and \citetalias{Lau2007} -- as well as for \citetalias{Kor2009} for comparison.

\section{An alternative, \emph{integral} characterization of early-type galaxies}
\label{sec: Delta3D}

It is apparent that measuring the logarithmic slopes of the inner regions of early-type galaxies can be a rather ambiguous process. In particular, the radius at which either $\gamma_{\rm 2D}$ or $\gamma_{\rm 3D}$ should be measured is not clear and is often not consistent from one study to the next. This is troubling given that, as we have shown in \S~\ref{sec: Results} and \S\ref{sec: Caveat}, $\gamma_{\rm 3D}$ can vary substantially over relatively small radii. 

In their paper addressing $\gamma_{\rm 2D}$, \citetalias{Cot2007} propose a new, physically motivated parameter designed to quantify the behavior of the inner surface brightness profile and the tendency of early-type galaxies to transition systematically from central light deficit below the global S\'{e}rsic fit for the brightest galaxies to a central light excess above the global S\'{e}rsic fit as galaxies become fainter. This quantity, dubbed $\Delta_{0.02}=\Delta_{\rm 2D}$, is essentially the logarithm of the ratio of the observed luminosity in the central region to the luminosity from the inward extrapolation of the S\'{e}rsic fit, both integrated within 2\% of the effective radius. For the deprojected profiles here, we define a similar quantity $\Delta_{\rm 3D}$ as:
\begin{equation}
\label{eq: Delta3D}
\Delta_{\rm 3D}=\log \left(\mathcal{L}_{\rm{gal}}/\mathcal{L}_{\rm{Ser}}\right),
\end{equation}
where $\mathcal{L}_{\rm{gal}}$ and $\mathcal{L}_{\rm{Ser}}$ are the luminosities inside the break radius $R_b$, integrated from the deprojected core-S\'{e}rsic profile for the former and from the deprojection of the S\'{e}rsic-only profile for the latter. This parameter allows us to quantitatively characterize a galaxy as having a light deficit ($\Delta_{\rm 3D}<0$) or a light excess (i.e., a nucleus; $\Delta_{\rm 3D}>0$). 

\begin{figure}
\includegraphics[width=6in]{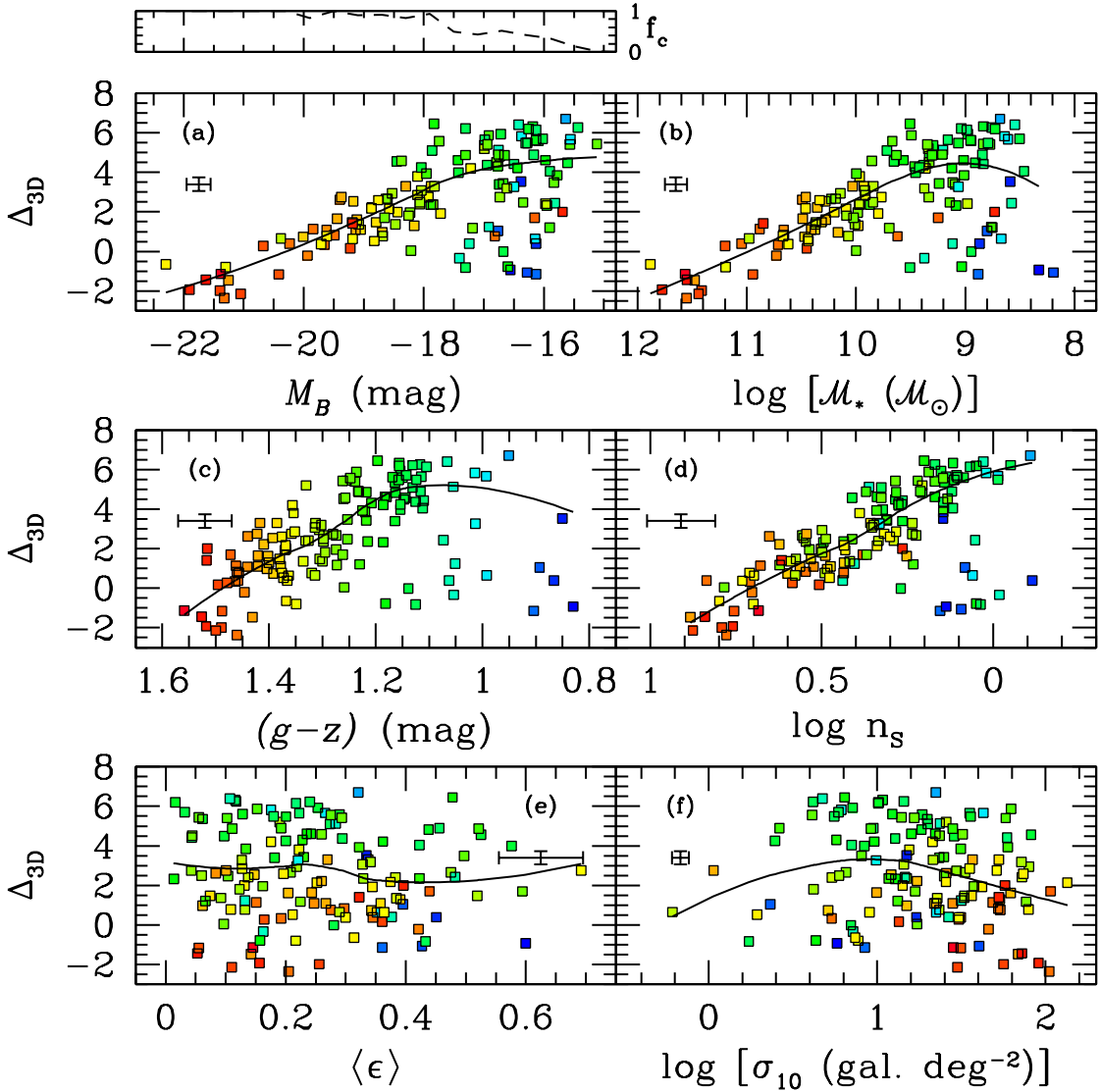}
\caption{Behavior of the parameter $\Delta_{\rm 3D}$ = log (${\cal L}_{{\rm gal}} /{\cal L}_{{\rm Ser}}$) from z-band photometry for  ACSVCS and ACSFCS galaxies. This parameter represents quantitatively whether a galaxy has a central luminosity deficit ($\Delta_{\rm 3D}$ $<$ 0) with respect to the S\'ersic fit to a galaxy, a central excess ($\Delta_{\rm 3D}$ $>$ 0), or neither ($\Delta_{\rm 3D}$ = 0). Panels $(a)$ to $(f)$ show the relationship between $\Delta_{\rm 3D}$ and absolute blue magnitude, galaxy (stellar) mass (from \citealt{Pen2008}), galaxy color, S\'ersic index, mean ellipticity, and the surrounding density of Virgo or Fornax cluster members. In all panels, galaxies have been color-coded according to their $(g-z)$ color, as shown in panel $(c)$. The smooth curve in each panel shows the locally-weighted scatterplot smoothing (LOWESS) fit to the data (see \citealt{Cle1984}) highlighting the general trends, if any; representative error bars are shown in each panel. For reference, the completeness fraction, $f_c$ , of our sample of early-type galaxies in the
Virgo and Fornax Clusters as a function of absolute magnitude is shown in the upper left panel. This figure illustrates the clear trend from light deficit to light excess as one moves down the galaxy luminosity/mass function, although there is a spread in $\Delta_{\rm 3D}$ for the faintest and lowest mass galaxies. In particular, note the presence of the faint ($M_B \gtrsim -18.5$~mag), weakly/non-nucleated galaxies
which represent $\sim$10\% of our sample.}
\label{fig:Delta}
\end{figure}

We have computed $\Delta_{\rm 3D}$ for each galaxy in our sample, the results of which are listed in Tables~\ref{tab:acsvcs} and \ref{tab:acsfcs}. In Figure~\ref{fig:Delta}, these values are plotted  against a variety of galaxy parameters (e.g., \citealt{Pen2008}; \citealt{Fer2006}; L.~Ferrarese et al. 2011, \emph{in prep}; P.~C\^ot\'e et~al. 2011, \emph{in prep}). Panels~$(a)$ -- $(f)$ show $\Delta_{\rm 3D}$ as a function of absolute blue magnitude~$M_B$, galaxy (stellar) mass~$\cal M_*$, $(g-z)$ color, S\'ersic index~$n=n_S$, mean ellipticity~$\langle\epsilon\rangle$, and local galaxy density~$\sigma_{10}$\footnote{Defined as the surface density corresponding
to the distance of the 10 nearest confirmed or probable cluster members according to \citet{Bin1985} and \citet{Fer1989}, i.e., $\sigma_{10} = {\pi}R_{10}^2$.}.
Symbols have been color-coded according to the galaxy $(g-z)$ colors, which are shown in panel $(c)$. To highlight any trends with $\Delta_{\rm 3D}$, the smooth curve shown in each panel shows the (non-parametric) locally-weighted scatterplot smoothing (LOWESS) fit to the data (see \citealt{Cle1984}). It is advisable to view the extremes of the LOWESS curves with some caution given their reliance on a small number of data points. In the small panel in the upper left we show the completeness fraction $f_c$ of our sample of early-type galaxies in the Virgo and Fornax Clusters as a function of absolute magnitude.

There are obvious correlations in a number of these panels: i.e., there is a systematic increase in $\Delta_{\rm 3D}$ as one moves to lower luminosities or masses, bluer colors
and lower S\'ersic indices. Note that although most intermediate- and low-luminosity galaxies in our sample are strongly nucleated, there is a population of faint ($M_B \gtrsim -18.5$), blue galaxies which either have smaller-than-expected nuclei or which are not nucleated at all, resulting in a skewed $\Delta_{3D}$ distribution around this magnitude. These could be galaxies that have not yet completed Ð or indeed begun Ð forming their nuclei. (See \S4 of \citetalias{Cot2007} for further discussion.) The most luminous, most massive, and reddest galaxies are always characterized by central light deficits, $\Delta_{\rm 3D} < 0$, although there is also a small number of faint and anomalously red galaxies that have $\Delta_{\rm 3D} > 0$. These are compact elliptical galaxies, which may be the tidal-stripped remains of more massive systems (e.g., \citealt{Fab1973}, \citealt{Bek2001}, \citealt{Chi2009b}, P.~C\^ot\'e et al. 2011 \emph{in prep}).  By contrast, there is little or no correlation between $\Delta_{\rm 3D}$ and mean ellipticity or galaxy density --- aside from the well known tendency for deficit/core galaxies to have low ellipticities in their central regions, and to occupy high-density environments at the centers of clusters or subclusters. 

There are many advantages to using $\Delta_{\rm 3D}$ over $\gamma_{\rm 3D}$ to characterize the inner regions of early-type galaxies. Because $\Delta_{\rm 3D}$ is an integral quantity, it is not measured at an instantaneous radius and is therefore much easier to measure consistently than $\gamma_{\rm 3D}$ (although, of course $\Delta_{\rm 3D}$ is not model-independent). The inner region is simply defined as the radius inside which the total galaxy profile departs from the global S\'{e}rsic profile. Note that this does not necessarily require fitting a core-S\'{e}rsic profile in particular to a given galaxy. One need only combine a good global fit (likely a S\'{e}rsic profile) with a fit to the central region. This is advantageous, for example, in the case of a distant galaxy with an unresolved nucleus: the nucleus could be simply modeled as a point source and a value for $\Delta_{\rm 3D}$ obtained, whereas it would not be possible to measure $\gamma_{\rm 3D}$ including the nucleus for such a galaxy. Additionally, $\Delta_{\rm 3D}$ indicates whether a galaxy is nucleated or non-nucleated in a clear, intuitive way; conversely, it is not possible to state whether a galaxy contains a nucleus or not simply by knowing $\gamma_{\rm 3D}$.

\section{Discussion}
\label{sec: Discussion}

The main result of this paper is that the deprojected profiles of the ACS Virgo and Fornax Cluster Survey galaxies do not support the existence of a core/power-law dichotomy around $M_B\sim-20.5$~mag. Rather, we find that the inner luminosity density profiles fan out over a continuum of slopes, as \citetalias{Cot2007} found when analyzing the \emph{projected} profiles of the same sample. This result holds whether the compact nuclear components (i.e., nuclei, present in the vast majority of galaxies fainter than $M_B \sim -19$ mag) are included or excluded in the deprojection. 

This finding is in contrast to the results of \citetalias{Lau2007} who also analyzed the projected and deprojected inner slopes of a sample of galaxies for which ``Nuker"-law fits to the surface brightness profiles were available in the literature. As discussed in \S \ref{sec: Caveat}, the actual details of the deprojection technique are unlikely to be responsible for the difference in our findings. \citetalias{Cot2007} point out that the \citetalias{Lau2007} results are, in fact, biased by their sample selection, which is described by a luminosity function (shown in Figure~\ref{fig:lumfunc}) that is itself bimodal, as \citetalias{Lau2007} themselves also note. Given the observed trend between galaxy magnitudes and inner profile slopes, \citetalias{Cot2007} show  that inner slopes drawn from a continuous distribution (such as the one observed for the ACSVCS/FCS galaxies) for galaxies following a bimodal luminosity function (such as the one characterizing the sample of  \citetalias{Lau2007}) will produce a bimodal slope distribution that closely resembles the one observed by \citetalias{Lau2007}. The analysis presented in this paper adds further weight to this explanation by showing that, when using a representative galaxy sample such as the ACSVCS/FCS (see Figure~\ref{fig:lumfunc}), both the two-dimensional surface brightness profiles and the three-dimensional deprojected profiles define a nearly continuous sequence as a function of galaxy magnitude. Indeed, as shown in Figure~\ref{fig:profiles}, the apparent contrast in central brightness profiles between the brightest (shallow slopes) and the fainter galaxies (steeper slopes) is less striking in the deprojected profiles, as noted by previous investigators (e.g., \citetalias{Geb1996}, \citetalias{Lau2007}).

The absence of a dichotomy between ``core'' and ``power-law'' galaxies should perhaps not  be surprising. It is generally believed that the core galaxies that populate the upper end of the luminosity function are nearly spherically symmetric, pressure supported, slowly rotating, boxy systems, while the opposite is true for power-law galaxies. In fact, although isophotal shape, kinematics and stellar populations do show systematic trends as a function of galaxy magnitude \citep{Ben1989, Kor1989, Cao1993, Fer2006}, there is {\it not} a one-to-one correspondence between the core/power-law classification and the above mentioned properties. Most notably, the SAURON team found no clear correspondence between ``core/power-law" galaxies and their ``slow/fast rotators" \citep{Ems2007}. Similarly, P.~C\^ot\'e et al. (2011, \emph{in prep}) find that stellar content and global structural parameters of ACSVCS and ACSFCS galaxies vary systematically along the luminosity function, but do not show any sign of a {\it discontinuity} across the alleged ``core/power-law" divide.
 
From a theoretical standpoint, a dichotomy is also difficult to reconcile with a hierarchical merging scenario for the formation of early-type galaxies. Based on hydrodynamic simulations, \citet{Hop2008, Hop2009} find that the ``core" and ``power-law" galaxies actually form a continuous family, very much in agreement with the results presented in this paper, as well as in \citet{Fer2006} and \citetalias{Cot2007}. A continuity is, in fact, expected given that the key processes involved in the formation of spheroids (e.g., the degree of dissipation) depend on factors such as the gas fractions and masses of the progenitors, which are themselves not believed to be discontinuous. This is not at all to argue that the faintest galaxies in our sample are simply scaled-down versions of the giant ellipticals. There are a number of processes that likely affect these galaxies, including mergers (with various gas fractions), stripping, harassment, cold gas accretion, etc. However, these processes should have differing -- but not \emph{discontinuous} -- levels of importance as we travel down the luminosity function. In our view, parsing complete, or nearly complete, galaxy samples into small subgroups where certain physical mechanisms are expected to dominate and then concluding that these populations are fundamentally distinct will lead to an overly simplified, not to mention logically cyclical, view of the galaxy formation process. 

As an aside, note that, moving down the luminosity function, our results also show no evidence of a discontinuity across the so called ``giant/dwarf'' transition, traditionally thought to occur at $M_B\approx-17.5$~mag (e.g., \citealt{Kor1985}). One of the arguments often cited in support of the notion that dwarf galaxies are physically distinct from regular ellipticals is that the former have exponential surface brightness profiles (i.e., S\'{e}rsic index $n\sim1$). Indeed, as Figure~\ref{fig:profiles}\emph{h-j} demonstrates, as one travels down to fainter galaxies (from VCC~828 to VCC~1075 in this case) the underlying galaxy at inner radii tends to become very flat in projection. This trend may make it tempting to conclude that galaxies with flat underling inner surface brightness profiles are dwarf ellipticals while those with steeper inner slopes are giants. However, as Figure~\ref{fig:gzcompare} illustrates, whereas $\gamma_{\rm 3D} \approx \gamma_{\rm 2D}+1$ for $\gamma_{\rm 2D} \gtrsim 1$, since $\gamma_{\rm 2D} \lesssim 1 $ tends to deproject to a range of values between $\gamma_{\rm 2D}$ and $\gamma_{\rm 2D}+1$, many of those ``dwarf'' galaxies that appear flat in projection fan out to create a continuous trend with magnitude when deprojected, bridging the gap to the so-called ``giants''. Referring back to Figure~\ref{fig:gammaMBnonuc}, it is apparent that $\gamma_{\rm 3D}$ of the underlying galaxy is continuous with magnitude and does not suddenly flatten around $M_B\approx-17.5$.

At this stage, it is worth emphasizing that although S\'{e}rsic profiles were used to parameterize the observed surface brightness profiles globally simply because they provided the best empirical match, the very fact that this family of models so accurately fits the surface brightness profiles (for $R \gtrsim 2\%R_e$) of both ``core" and ``power-law" galaxies, and of high- (``giant") and low-luminosity (``dwarf") galaxies must be a fundamental clue to the physics underlying the hierarchical assembly of baryons within merging dark matter halos. So is there any evidence for a physically motivated origin for S\'{e}rsic model? \citet{Hjo1995} were the first to show that the deviations of the brightness profiles of real galaxies from a de Vaucouleurs $R^{1/4}$ law --- the same deviations that can be accounted for explicitly by a S\'{e}rsic law --- can be reproduced using a simple distribution function constructed on the basis of the statistical mechanics of violent relaxation. In a series of papers by \citet{Ger1997}, \citet{Lim1999}, \citet{Mar2000}, and \citet{Mar2001} it has been shown that elliptical galaxies lie on the intersection between two manifolds --- one a scaling relation between potential energy and mass and the other representing quasi-constant specific entropy. These investigators were able to express this intersection in terms of the three S\'{e}rsic parameters and verify that actual elliptical galaxies fall within the S\'{e}rsic parameter space predicted theoretically. Finally, in a review article, \citet{Lon2006} discusses how N-body numerical simulations of collisionless gravitating systems can reproduce S\'ersic profiles given appropriate initial conditions. 

Finally, in terms of the origin of galaxies with central light deficits versus those with light excess, it is widely accepted that the light deficits in ``core" galaxies are the result of central scouring by coalescing black hole binaries following predominantly dissipationless galaxy mergers (\citealt{Ebi1991, Fab1997, Gua2008}) whereas nuclei are thought to be mainly the result of gas inflows into the core (\citealt{Mih1994, Cot2006}; \citetalias{Cot2007}; \citealt{Ems2008, Kor2009}), with possible contributions from other processes, such as the infall of star clusters via dynamical friction (\citealt{Tre1975, Tre1976}). It has been suggested that gas inflows played a role in the centers of light-deficit galaxies as well (see, e.g., \citealt{Hop2009b}), although \citet{Ems2008} find that the tidal forces in a galaxy are compressive for S\'{e}rsic indices $n \lesssim 3.5$ such that available gas could collapse and form a cluster of stars in the center while, for $n \gtrsim 3.5$,  the tidal forces become disruptive. It may be that, although gas compression becomes increasingly important as one moves down the luminosity function, inflows still occur at higher masses, albeit at reduced levels.

\acknowledgments

The authors would like to thank the anonymous referee for suggestions which improved the manuscript. LG thanks Thomas Puzia and Chien Peng for all of their help. Support for programs GO-9401 and GO-10217 was provided through grants from STScI, which is operated by AURA, Inc., under NASA contract NAS5-26555. The authors gratefully acknowledge support from NSERC though the Discovery and Postgraduate Scholarship programs, as well as from the University of Victoria through their fellowship program.

\clearpage 
\begin{deluxetable}{clcccccccccccc} 
\tabletypesize{\scriptsize} 
\tablecolumns{14}  
\tablecaption{$\gamma_{\rm 3D}$ and $\Delta_{\rm 3D}$ values computed for ACS Virgo Cluster Survey galaxies \label{tab:acsvcs}} 
\tablewidth{0pt} 
\tablehead{ 
\colhead{ID} & \colhead{Name} & \colhead{$M_B$} & \multicolumn{2}{c}{$\gamma_{\rm 3D}(0.5\%R_e)$} & \multicolumn{2}{c}{$\gamma_{\rm 3D}(1\%R_e)$} & \multicolumn{2}{c}{$\gamma_{\rm 3D}(5\%R_e)$} & \multicolumn{2}{c}{$\gamma_{\rm 3D}(30\%R_e)$} & \multicolumn{2}{c}{$\gamma_{\rm 3D}(0\Sec1)$} & \colhead{$\Delta_{\rm 3D}$} \\ 
\colhead{(1)} & \colhead{(2)} & \colhead{(3)} & \multicolumn{2}{c}{(4)} & \multicolumn{2}{c}{(5)} & \multicolumn{2}{c}{(6)} & \multicolumn{2}{c}{(7)} & \multicolumn{2}{c}{(8)} & \colhead{(9)} \\ 
} 
\startdata 
1 & VCC 1226 & -21.90 & 0.30 & \nodata & 1.00 & \nodata & 2.02 & \nodata & 2.46 & \nodata & 0.70 & \nodata & -1.99 \\ 
  &  &  & 0.32 & \nodata & 1.08 & \nodata & 2.04 & \nodata & 2.47 & \nodata & 0.76 & \nodata & -1.93 \\ 
2 & VCC 1316 & -21.63 & 1.09 & \nodata & 0.93 & \nodata & 1.38 & \nodata & 2.52 & \nodata & 1.23 & \nodata & -1.74 \\ 
  &  &  & 1.15 & \nodata & 1.01 & \nodata & 1.33 & \nodata & 2.49 & \nodata & 1.25 & \nodata & -1.43 \\ 
3 & VCC 1978 & -21.38 & 0.95 & \nodata & 0.69 & \nodata & 1.78 & \nodata & 2.37 & \nodata & 1.15 & \nodata & -1.04 \\ 
  &  &  & 0.95 & \nodata & 0.69 & \nodata & 1.80 & \nodata & 2.40 & \nodata & 1.15 & \nodata & -1.14 \\ 
4 & VCC 881 & -21.33 & 1.69 & \nodata & 1.79 & \nodata & 2.07 & \nodata & 2.48 & \nodata & 0.30 & \nodata & -2.26 \\ 
  &  &  & 1.69 & \nodata & 1.79 & \nodata & 2.07 & \nodata & 2.48 & \nodata & 0.25 & \nodata & -2.36 \\ 
5 & VCC 798 & -21.30 & 0.53 & \nodata & 1.37 & \nodata & 1.93 & \nodata & 2.39 & \nodata & 0.96 & \nodata & -0.90 \\ 
  &  &  & 0.53 & \nodata & 1.52 & \nodata & 1.93 & \nodata & 2.39 & \nodata & 1.01 & \nodata & -0.79 \\ 
6 & VCC 763 & -21.25 & 0.86 & \nodata & 0.76 & \nodata & 2.23 & \nodata & 2.59 & \nodata & 1.25 & \nodata & -1.41 \\ 
  &  &  & 0.89 & \nodata & 0.74 & \nodata & 2.27 & \nodata & 2.61 & \nodata & 1.27 & \nodata & -1.47 \\ 
7 & VCC 731 & -21.40 & 0.71 & \nodata & 0.78 & \nodata & 2.07 & \nodata & 2.50 & \nodata & 1.12 & \nodata & -1.31 \\ 
  &  &  & 0.66 & \nodata & 1.29 & \nodata & 2.22 & \nodata & 2.59 & \nodata & 0.95 & \nodata & -1.98 \\ 
8 & VCC 1535 & -20.58 & \nodata & \nodata & \nodata & \nodata & \nodata & \nodata & \nodata & \nodata & \nodata & \nodata & \nodata \\ 
  &   &   & \nodata & \nodata & \nodata & \nodata & \nodata & \nodata & \nodata & \nodata & \nodata & \nodata & \nodata \\ 
9 & VCC 1903 & -20.24 & \nodata & \nodata & \nodata & \nodata & \nodata & \nodata & \nodata & \nodata & \nodata & \nodata & \nodata \\ 
  &  &  & 1.93 & 1.90 & 1.86 & 2.00 & 2.24 & 2.25 & 2.59 & 2.59 & 1.95 & 1.69 & 0.39 \\ 
10 & VCC 1632 & -20.42 & 0.43 & \nodata & 1.38 & \nodata & 2.08 & \nodata & 2.49 & \nodata & 0.95 & \nodata & -1.20 \\ 
  &  &  & 0.49 & \nodata & 1.27 & \nodata & 2.08 & \nodata & 2.49 & \nodata & 0.98 & \nodata & -1.16 \\ 
11 & VCC 1231 & -19.94 & 0.81 & \nodata & 0.89 & \nodata & 1.65 & \nodata & 2.21 & \nodata & 0.95 & \nodata & -0.19 \\ 
  &  &  & 0.82 & \nodata & 0.86 & \nodata & 1.65 & \nodata & 2.21 & \nodata & 0.95 & \nodata & -0.21 \\ 
12 & VCC 2095 & -20.01 & 1.45 & 1.41 & 1.41 & 1.51 & 1.59 & 1.82 & 2.29 & 2.32 & 1.44 & 1.43 & 0.05 \\ 
  &  &  & 1.50 & 1.23 & 1.43 & 1.33 & 1.62 & 1.64 & 2.19 & 2.19 & 1.46 & 1.30 & 0.74 \\ 
13 & VCC 1154 & -19.85 & 1.42 & 1.64 & 1.63 & 1.74 & 2.03 & 2.03 & 2.46 & 2.46 & 1.51 & 1.56 & 0.25 \\ 
  &  &  & 1.67 & 1.33 & 1.52 & 1.43 & 1.74 & 1.74 & 2.26 & 2.26 & 1.67 & 1.25 & 1.14 \\ 
14 & VCC 1062 & -19.62 & 1.62 & 1.07 & 1.61 & 1.17 & 1.47 & 1.47 & 2.06 & 2.06 & 1.62 & 1.09 & 1.18 \\ 
  &  &  & 1.60 & 1.14 & 1.58 & 1.24 & 1.55 & 1.55 & 2.12 & 2.12 & 1.60 & 1.16 & 1.09 \\ 
15 & VCC 2092 & -19.66 & 1.80 & 1.38 & 1.50 & 1.48 & 1.79 & 1.79 & 2.29 & 2.29 & 1.79 & 1.35 & 1.27 \\ 
  &  &  & 1.58 & 1.60 & 1.64 & 1.70 & 1.94 & 1.99 & 2.42 & 2.43 & 1.56 & 1.53 & 0.34 \\ 
16 & VCC 369 & -19.40 & 2.02 & 0.81 & 2.01 & 0.89 & 1.63 & 1.17 & 1.81 & 1.80 & 1.98 & 0.94 & 2.80 \\ 
  &  &  & 1.89 & 0.76 & 1.86 & 0.83 & 1.66 & 1.11 & 1.82 & 1.74 & 1.83 & 0.89 & 2.62 \\ 
17 & VCC 759 & -19.49 & 1.41 & 1.23 & 1.26 & 1.33 & 1.64 & 1.64 & 2.19 & 2.19 & 1.44 & 1.15 & 0.69 \\ 
  &  &  & 1.50 & 1.27 & 1.35 & 1.37 & 1.69 & 1.69 & 2.22 & 2.22 & 1.52 & 1.19 & 0.83 \\ 
18 & VCC 1692 & -19.44 & 1.77 & 1.09 & 1.76 & 1.18 & 1.67 & 1.49 & 2.06 & 2.07 & 1.76 & 1.13 & 1.47 \\ 
  &  &  & 1.80 & 1.08 & 1.77 & 1.17 & 1.70 & 1.48 & 2.06 & 2.06 & 1.78 & 1.13 & 1.65 \\ 
19 & VCC 1030 & -19.39 & \nodata & \nodata & \nodata & \nodata & \nodata & \nodata & \nodata & \nodata & \nodata & \nodata & \nodata \\ 
  &   &   & \nodata & \nodata & \nodata & \nodata & \nodata & \nodata & \nodata & \nodata & \nodata & \nodata & \nodata \\ 
20 & VCC 2000 & -19.08 & \nodata & \nodata & \nodata & \nodata & \nodata & \nodata & \nodata & \nodata & \nodata & \nodata & \nodata \\ 
  &  &  & 1.90 & 1.35 & 1.91 & 1.45 & 1.76 & 1.76 & 2.27 & 2.27 & 1.90 & 1.44 & 1.36 \\ 
21 & VCC 685 & -19.22 & 1.41 & 1.29 & 1.37 & 1.39 & 1.56 & 1.70 & 2.24 & 2.24 & 1.41 & 1.30 & 0.09 \\ 
  &  &  & 1.45 & 1.27 & 1.42 & 1.37 & 0.57 & 1.68 & 2.22 & 2.22 & 1.44 & 1.29 & 0.17 \\ 
22 & VCC 1664 & -19.14 & 1.86 & 1.23 & 1.85 & 1.33 & 1.77 & 1.64 & 2.19 & 2.19 & 1.85 & 1.26 & 1.27 \\ 
  &  &  & 1.82 & 1.37 & 1.81 & 1.47 & 1.78 & 1.78 & 2.29 & 2.29 & 1.82 & 1.39 & 1.10 \\ 
23 & VCC 654 & -19.17 & 1.81 & 0.88 & 1.89 & 0.96 & 1.31 & 1.25 & 1.87 & 1.87 & 1.93 & 1.00 & 2.20 \\ 
  &  &  & 1.94 & 0.86 & 2.12 & 0.94 & 1.24 & 1.23 & 1.85 & 1.85 & 2.23 & 0.98 & 2.54 \\ 
24 & VCC 944 & -19.05 & 1.61 & 1.16 & 1.59 & 1.25 & 1.62 & 1.56 & 2.10 & 2.13 & 1.60 & 1.21 & 1.08 \\ 
  &  &  & 1.63 & 1.16 & 1.62 & 1.25 & 1.65 & 1.56 & 2.10 & 2.13 & 1.62 & 1.22 & 1.10 \\ 
25 & VCC 1938 & -19.19 & 1.91 & 1.27 & 1.91 & 1.37 & 1.68 & 1.68 & 2.22 & 2.22 & 1.91 & 1.31 & 1.51 \\ 
  &  &  & 1.92 & 1.28 & 1.92 & 1.37 & 1.70 & 1.69 & 2.22 & 2.22 & 1.92 & 1.34 & 1.48 \\ 
26 & VCC 1279 & -19.11 & 1.63 & 0.61 & 1.62 & 0.67 & 1.56 & 0.92 & 1.80 & 1.55 & 1.62 & 0.64 & 2.06 \\ 
  &  &  & 1.64 & 0.77 & 1.63 & 0.85 & 1.59 & 1.12 & 1.84 & 1.75 & 1.63 & 0.82 & 1.62 \\ 
27 & VCC 1720 & -18.87 & 2.03 & 0.86 & 2.04 & 0.94 & 1.87 & 1.23 & 1.87 & 1.85 & 2.04 & 0.94 & 2.60 \\ 
  &  &  & 1.97 & 0.89 & 1.97 & 0.98 & 1.96 & 1.27 & 1.89 & 1.89 & 1.98 & 0.98 & 2.22 \\ 
28 & VCC 355 & -18.70 & 1.71 & 1.00 & 1.70 & 1.09 & 1.40 & 1.39 & 1.99 & 1.99 & 1.70 & 1.15 & 1.50 \\ 
  &  &  & 1.55 & 1.05 & 1.53 & 1.14 & 1.48 & 1.45 & 2.02 & 2.04 & 1.51 & 1.22 & 0.97 \\ 
29 & VCC 1619 & -18.60 & 1.81 & 0.56 & 1.81 & 0.62 & 1.87 & 0.85 & 1.48 & 1.48 & 1.81 & 0.61 & 2.56 \\ 
  &  &  & 1.86 & 0.60 & 1.87 & 0.67 & 2.03 & 0.91 & 1.54 & 1.54 & 1.87 & 0.65 & 2.67 \\ 
30 & VCC 1883 & -18.63 & 2.28 & 0.78 & 2.33 & 0.85 & 1.76 & 1.13 & 1.76 & 1.76 & 2.44 & 0.95 & 3.48 \\ 
  &  &  & 2.01 & 0.80 & 2.00 & 0.87 & 2.83 & 1.15 & 1.79 & 1.79 & 2.03 & 0.97 & 2.66 \\ 
31 & VCC 1242 & -18.53 & 1.91 & 0.67 & 1.93 & 0.73 & 1.65 & 0.99 & 1.62 & 1.62 & 1.95 & 0.76 & 2.81 \\ 
  &  &  & 1.95 & 0.66 & 1.97 & 0.72 & 1.71 & 0.98 & 1.61 & 1.61 & 2.01 & 0.76 & 2.94 \\ 
32 & VCC 784 & -18.44 & 2.18 & 1.21 & 1.60 & 1.31 & 1.62 & 1.62 & 2.17 & 2.17 & 2.29 & 1.23 & 2.17 \\ 
  &  &  & 2.08 & 1.26 & 1.51 & 1.36 & 1.67 & 1.67 & 2.21 & 2.21 & 2.12 & 1.28 & 1.99 \\ 
33 & VCC 1537 & -18.47 & 1.70 & 0.95 & 1.70 & 1.04 & 1.65 & 1.34 & 1.95 & 1.95 & 1.70 & 1.06 & 1.40 \\ 
  &  &  & 1.72 & 1.00 & 1.72 & 1.09 & 1.65 & 1.39 & 1.99 & 1.99 & 1.72 & 1.11 & 1.37 \\ 
34 & VCC 778 & -18.68 & 1.53 & 1.10 & 1.51 & 1.19 & 1.20 & 1.50 & 2.08 & 2.08 & 1.48 & 1.27 & 0.79 \\ 
  &  &  & 1.53 & 1.09 & 1.52 & 1.19 & 1.14 & 1.50 & 2.08 & 2.08 & 1.48 & 1.29 & 0.77 \\ 
35 & VCC 1321 & -18.22 & 1.76 & 1.76 & 1.86 & 1.86 & 2.13 & 2.13 & 2.52 & 2.52 & 1.38 & 1.66 & 0.08 \\ 
  &  &  & 1.54 & 1.71 & 1.82 & 1.82 & 2.09 & 2.09 & 2.50 & 2.50 & 1.34 & 1.65 & 0.04 \\ 
36 & VCC 828 & -18.58 & 1.76 & 0.80 & 1.77 & 0.87 & 1.64 & 1.15 & 1.78 & 1.78 & 1.76 & 0.85 & 2.14 \\ 
  &  &  & 1.77 & 0.82 & 1.79 & 0.89 & 1.62 & 1.17 & 1.80 & 1.80 & 1.78 & 0.88 & 2.15 \\ 
37 & VCC 1250 & -18.45 & 2.17 & 0.61 & 2.25 & 0.67 & 0.91 & 0.91 & 1.54 & 1.54 & 2.39 & 0.70 & 3.81 \\ 
  &  &  & 2.47 & 0.60 & 2.66 & 0.66 & 0.91 & 0.91 & 1.54 & 1.54 & 3.32 & 0.70 & 4.56 \\ 
38 & VCC 1630 & -18.30 & 1.82 & 0.66 & 1.80 & 0.72 & 1.69 & 0.97 & 1.85 & 1.61 & 1.81 & 0.68 & 2.63 \\ 
  &  &  & 1.83 & 0.53 & 1.80 & 0.59 & 1.71 & 0.82 & 1.96 & 1.44 & 1.83 & 0.54 & 3.30 \\ 
39 & VCC 1146 & -18.23 & 2.16 & 1.14 & 2.16 & 1.23 & 2.15 & 1.54 & 2.12 & 2.12 & 2.16 & 1.12 & 2.18 \\ 
  &  &  & 2.16 & 1.24 & 2.16 & 1.34 & 2.15 & 1.65 & 2.20 & 2.20 & 2.16 & 1.22 & 1.96 \\ 
40 & VCC 1025 & -18.79 & 1.76 & 1.23 & 1.76 & 1.33 & 1.67 & 1.64 & 2.19 & 2.19 & 1.76 & 1.31 & 1.34 \\ 
  &  &  & 1.83 & 1.23 & 1.83 & 1.33 & 1.71 & 1.64 & 2.19 & 2.19 & 1.83 & 1.33 & 1.48 \\ 
41 & VCC 1303 & -18.11 & 2.02 & 0.86 & 1.95 & 0.94 & 1.81 & 1.23 & 2.05 & 1.85 & 1.99 & 0.89 & 3.55 \\ 
  &  &  & 2.02 & 0.81 & 1.95 & 0.89 & 1.84 & 1.17 & 2.09 & 1.79 & 1.99 & 0.84 & 3.80 \\ 
42 & VCC 1913 & -18.07 & 1.98 & 0.93 & 2.01 & 1.02 & 1.54 & 1.31 & 1.93 & 1.93 & 1.99 & 0.96 & 2.50 \\ 
  &  &  & 1.98 & 0.98 & 2.01 & 1.07 & 1.58 & 1.37 & 1.97 & 1.97 & 1.99 & 1.01 & 2.40 \\ 
43 & VCC 1327 & -18.16 & 1.41 & 0.97 & 1.38 & 1.06 & 1.36 & 1.36 & 1.96 & 1.96 & 1.35 & 1.09 & 1.09 \\ 
  &  &  & 1.56 & 0.94 & 1.55 & 1.02 & 1.32 & 1.32 & 1.93 & 1.93 & 1.54 & 1.08 & 1.30 \\ 
44 & VCC 1125 & -17.92 & 1.95 & 0.84 & 2.47 & 0.92 & 1.20 & 1.20 & 1.83 & 1.83 & 2.01 & 0.86 & 2.67 \\ 
  &  &  & 1.84 & 0.70 & 1.98 & 0.76 & 1.02 & 1.02 & 1.66 & 1.66 & 1.92 & 0.75 & 2.75 \\ 
45 & VCC 1475 & -17.89 & 1.88 & 1.10 & 1.95 & 1.20 & 1.50 & 1.50 & 2.09 & 2.09 & 1.94 & 1.19 & 1.94 \\ 
  &  &  & 2.15 & 1.16 & 2.81 & 1.26 & 1.57 & 1.57 & 2.13 & 2.13 & 3.22 & 1.25 & 2.36 \\ 
46 & VCC 1178 & -17.72 & 1.89 & 1.14 & 1.87 & 1.24 & 1.81 & 1.55 & 2.14 & 2.12 & 1.84 & 1.31 & 1.86 \\ 
  &  &  & 1.85 & 1.09 & 1.83 & 1.18 & 1.81 & 1.49 & 2.14 & 2.08 & 1.81 & 1.26 & 1.92 \\ 
47 & VCC 1283 & -17.88 & 2.29 & 0.83 & 2.39 & 0.91 & 1.20 & 1.19 & 1.82 & 1.82 & 2.32 & 0.84 & 3.32 \\ 
  &  &  & 2.46 & 0.91 & 2.21 & 0.99 & 1.28 & 1.28 & 1.90 & 1.90 & 2.44 & 0.90 & 3.30 \\ 
48 & VCC 1261 & -17.86 & 2.79 & 0.70 & 2.83 & 0.77 & 1.17 & 1.03 & 1.74 & 1.68 & 2.78 & 0.69 & 5.16 \\ 
  &  &  & 2.86 & 0.70 & 2.85 & 0.77 & 1.15 & 1.04 & 1.73 & 1.68 & 2.86 & 0.70 & 5.28 \\ 
49 & VCC 698 & -17.85 & 2.08 & 1.16 & 1.91 & 1.26 & 1.61 & 1.57 & 2.14 & 2.14 & 2.06 & 1.19 & 2.47 \\ 
  &  &  & 2.13 & 1.18 & 2.69 & 1.28 & 1.59 & 1.59 & 2.15 & 2.15 & 2.22 & 1.21 & 2.36 \\ 
50 & VCC 1422 & -17.43 & 2.63 & 0.69 & 3.61 & 0.76 & 1.02 & 1.02 & 1.65 & 1.65 & 2.62 & 0.69 & 4.66 \\ 
  &  &  & 2.73 & 0.76 & 5.80 & 0.84 & 1.11 & 1.11 & 1.74 & 1.74 & 2.68 & 0.75 & 4.61 \\ 
51 & VCC 2048 & -17.42 & 2.27 & 0.63 & 3.02 & 0.70 & 0.95 & 0.95 & 1.58 & 1.58 & 2.70 & 0.69 & 3.94 \\ 
  &  &  & 2.29 & 0.63 & 3.07 & 0.69 & 0.94 & 0.94 & 1.57 & 1.57 & 2.77 & 0.68 & 3.98 \\ 
52 & VCC 1871 & -17.22 & 2.43 & 0.55 & 2.33 & 0.61 & 1.84 & 0.84 & 1.66 & 1.46 & 2.26 & 0.64 & 4.94 \\ 
  &  &  & 2.41 & 0.68 & 2.35 & 0.75 & 1.78 & 1.01 & 1.71 & 1.64 & 2.29 & 0.79 & 4.20 \\ 
53 & VCC 9 & -17.41 & 0.89 & \nodata & 0.77 & \nodata & 0.29 & \nodata & 1.38 & \nodata & 0.98 & \nodata & -0.13 \\ 
  &  &  & 0.82 & \nodata & 0.67 & \nodata & 0.14 & \nodata & 1.49 & \nodata & 0.96 & \nodata & -0.34 \\ 
54 & VCC 575 & -17.68 & \nodata & \nodata & \nodata & \nodata & \nodata & \nodata & \nodata & \nodata & \nodata & \nodata & \nodata \\ 
  &   &   & \nodata & \nodata & \nodata & \nodata & \nodata & \nodata & \nodata & \nodata & \nodata & \nodata & \nodata \\ 
55 & VCC 1910 & -16.99 & 2.69 & 0.54 & 2.87 & 0.59 & 0.95 & 0.82 & 1.44 & 1.44 & 2.77 & 0.57 & 5.28 \\ 
  &  &  & 2.64 & 0.53 & 2.84 & 0.58 & 0.86 & 0.80 & 1.43 & 1.43 & 2.74 & 0.56 & 5.20 \\ 
56 & VCC 1049 & -16.92 & 1.37 & 0.89 & 1.33 & 0.97 & 1.26 & 1.26 & 1.88 & 1.88 & 1.34 & 0.96 & 1.11 \\ 
  &  &  & 1.30 & 0.70 & 1.28 & 0.77 & 1.03 & 1.03 & 1.66 & 1.66 & 1.27 & 0.77 & 1.24 \\ 
57 & VCC 856 & -16.99 & 2.49 & 0.20 & 2.51 & 0.23 & 2.52 & 0.38 & 0.99 & 0.89 & 2.50 & 0.21 & 5.65 \\ 
  &  &  & 2.50 & 0.33 & 2.53 & 0.37 & 2.32 & 0.55 & 1.17 & 1.13 & 2.51 & 0.34 & 5.36 \\ 
58 & VCC 140 & -16.94 & 2.29 & 0.44 & 3.11 & 0.49 & 0.70 & 0.69 & 1.30 & 1.30 & 3.23 & 0.49 & 4.39 \\ 
  &  &  & 2.33 & 0.46 & 3.94 & 0.51 & 0.71 & 0.71 & 1.33 & 1.33 & 4.16 & 0.51 & 4.37 \\ 
59 & VCC 1355 & -16.98 & 2.85 & 0.43 & 3.43 & 0.48 & 0.71 & 0.68 & 1.29 & 1.29 & 2.79 & 0.42 & 5.99 \\ 
  &  &  & 3.00 & 0.56 & 4.13 & 0.62 & 0.85 & 0.85 & 1.48 & 1.48 & 2.78 & 0.53 & 5.65 \\ 
60 & VCC 1087 & -16.92 & 2.96 & 0.47 & 3.64 & 0.52 & 0.74 & 0.73 & 1.34 & 1.34 & 3.01 & 0.47 & 6.13 \\ 
  &  &  & 2.85 & 0.49 & 3.44 & 0.55 & 0.76 & 0.76 & 1.38 & 1.38 & 2.88 & 0.50 & 5.87 \\ 
61 & VCC 1297 & -16.82 & 1.31 & 1.00 & 1.26 & 1.09 & 1.28 & 1.39 & 1.99 & 1.99 & 0.83 & 1.30 & 0.43 \\ 
  &  &  & 1.44 & 1.00 & 1.41 & 1.09 & 1.32 & 1.39 & 1.99 & 1.99 & 1.14 & 1.30 & 0.76 \\ 
62 & VCC 1861 & -16.79 & 2.21 & 0.34 & 2.23 & 0.38 & 2.52 & 0.56 & 1.14 & 1.14 & 2.22 & 0.37 & 4.53 \\ 
  &  &  & 2.20 & 0.34 & 2.22 & 0.38 & 1.01 & 0.56 & 1.15 & 1.15 & 2.21 & 0.37 & 4.52 \\ 
63 & VCC 543 & -16.73 & 2.14 & 0.43 & 6.66 & 0.48 & 0.68 & 0.68 & 1.29 & 1.29 & 2.66 & 0.45 & 3.83 \\ 
  &  &  & 4.18 & 0.43 & 0.48 & 0.48 & 0.68 & 0.68 & 1.29 & 1.29 & 1.45 & 0.46 & 4.82 \\ 
64 & VCC 1431 & -16.73 & 1.84 & 0.45 & 1.85 & 0.50 & 2.01 & 0.71 & 1.33 & 1.32 & 1.85 & 0.50 & 3.05 \\ 
  &  &  & 1.77 & 0.55 & 1.78 & 0.60 & 2.12 & 0.83 & 1.46 & 1.46 & 1.78 & 0.60 & 2.75 \\ 
65 & VCC 1528 & -16.68 & 2.25 & 0.59 & 2.84 & 0.65 & 0.89 & 0.89 & 1.52 & 1.52 & 5.86 & 0.68 & 3.96 \\ 
  &  &  & 2.03 & 0.58 & 2.46 & 0.64 & 0.88 & 0.88 & 1.51 & 1.51 & 5.12 & 0.68 & 3.44 \\ 
66 & VCC 1695 & -16.76 & 2.43 & 0.47 & 2.99 & 0.52 & 0.74 & 0.74 & 1.35 & 1.35 & 3.81 & 0.54 & 4.75 \\ 
  &  &  & 2.56 & 0.45 & 3.20 & 0.50 & 0.70 & 0.70 & 1.31 & 1.31 & 4.33 & 0.51 & 5.13 \\ 
67 & VCC 1833 & -16.67 & 1.02 & 0.75 & 0.90 & 0.81 & 0.67 & 1.08 & 1.71 & 1.72 & 0.84 & 0.84 & 0.26 \\ 
  &  &  & 0.92 & 0.92 & 0.77 & 0.77 & 0.66 & 0.66 & 1.73 & 1.73 & 0.72 & 0.72 & -0.01 \\ 
68 & VCC 437 & -16.77 & 2.46 & 0.69 & 2.69 & 0.75 & 1.01 & 1.01 & 1.65 & 1.65 & 2.45 & 0.67 & 4.46 \\ 
  &  &  & 2.44 & 0.72 & 2.76 & 0.79 & 1.05 & 1.05 & 1.69 & 1.69 & 2.42 & 0.70 & 4.30 \\ 
69 & VCC 2019 & -16.72 & 2.90 & 0.43 & 2.99 & 0.47 & 1.14 & 0.67 & 1.28 & 1.28 & 2.93 & 0.45 & 6.18 \\ 
  &  &  & 3.05 & 0.59 & 3.20 & 0.64 & 0.93 & 0.88 & 1.51 & 1.51 & 3.06 & 0.59 & 5.93 \\ 
70 & VCC 33 & -16.38 & 2.60 & 0.24 & 2.96 & 0.27 & 0.63 & 0.42 & 0.96 & 0.96 & 3.05 & 0.28 & 5.84 \\ 
  &  &  & 2.64 & 0.24 & 3.06 & 0.28 & 0.60 & 0.43 & 0.97 & 0.97 & 3.14 & 0.28 & 5.83 \\ 
71 & VCC 200 & -16.75 & 2.35 & 0.65 & 2.17 & 0.71 & 1.00 & 0.97 & 1.60 & 1.60 & 2.35 & 0.69 & 3.82 \\ 
  &  &  & 2.29 & 0.70 & 2.65 & 0.76 & 1.02 & 1.02 & 1.66 & 1.66 & 2.80 & 0.73 & 3.39 \\ 
72 & VCC 571 & -17.24 & 0.89 & 1.04 & 0.76 & 1.13 & 1.44 & 1.44 & 2.03 & 2.03 & 0.86 & 1.05 & 0.36 \\ 
  &  &  & 0.93 & 1.03 & 0.34 & 1.12 & 1.43 & 1.43 & 2.02 & 2.02 & 0.91 & 1.03 & 0.40 \\ 
73 & VCC 21 & -16.76 & 0.93 & 0.34 & 0.81 & 0.39 & 0.54 & 0.56 & 1.15 & 1.15 & 1.00 & 0.31 & 1.17 \\ 
  &  &  & 0.88 & 0.41 & 0.72 & 0.46 & 0.60 & 0.65 & 1.25 & 1.25 & 0.99 & 0.37 & 1.04 \\ 
74 & VCC 1488 & -16.38 & 2.26 & 0.25 & 2.92 & 0.29 & 0.44 & 0.44 & 0.99 & 0.99 & 4.62 & 0.30 & 4.94 \\ 
  &  &  & 1.65 & 0.23 & 1.75 & 0.27 & 0.42 & 0.42 & 0.95 & 0.95 & 1.79 & 0.27 & 3.53 \\ 
75 & VCC 1779 & -16.13 & -0.02 & \nodata & -0.03 & \nodata & 0.40 & \nodata & 1.53 & \nodata & -0.02 & \nodata & -1.37 \\ 
  &  &  & -0.01 & \nodata & -0.03 & \nodata & 0.33 & \nodata & 1.38 & \nodata & -0.02 & \nodata & -1.14 \\ 
76 & VCC 1895 & -16.17 & 2.34 & 0.30 & 3.34 & 0.34 & 0.52 & 0.51 & 1.08 & 1.08 & 3.79 & 0.35 & 4.87 \\ 
  &  &  & 2.36 & 0.29 & 3.52 & 0.32 & 0.49 & 0.49 & 1.05 & 1.05 & 3.94 & 0.34 & 4.89 \\ 
77 & VCC 1499 & -16.28 & -0.00 & \nodata & -0.01 & \nodata & -0.19 & \nodata & 1.33 & \nodata & -0.01 & \nodata & -1.28 \\ 
  &  &  & -0.00 & \nodata & -0.01 & \nodata & -0.27 & \nodata & 1.22 & \nodata & -0.01 & \nodata & -1.07 \\ 
78 & VCC 1545 & -16.34 & 1.95 & 0.92 & 2.18 & 1.01 & 1.31 & 1.31 & 1.92 & 1.92 & 2.09 & 0.99 & 2.50 \\ 
  &  &  & 2.01 & 0.98 & 2.63 & 1.07 & 1.37 & 1.37 & 1.97 & 1.97 & 2.37 & 1.06 & 2.48 \\ 
79 & VCC 1192 & -16.15 & 2.00 & 1.05 & 1.99 & 1.14 & 2.01 & 1.44 & 2.03 & 2.03 & 2.00 & 1.17 & 1.90 \\ 
  &  &  & 1.92 & 1.04 & 1.92 & 1.13 & 1.91 & 1.44 & 2.03 & 2.03 & 1.92 & 1.17 & 1.70 \\ 
80 & VCC 1857 & -16.13 & 1.00 & 0.15 & 0.94 & 0.18 & 0.61 & 0.31 & 0.31 & 0.79 & 1.02 & 0.14 & 0.66 \\ 
  &  &  & 1.07 & 0.30 & 1.03 & 0.34 & 0.79 & 0.50 & 0.20 & 1.07 & 1.08 & 0.28 & 0.39 \\ 
81 & VCC 1075 & -16.09 & 2.75 & 0.34 & 2.98 & 0.38 & 0.69 & 0.56 & 1.14 & 1.14 & 2.80 & 0.36 & 6.07 \\ 
  &  &  & 2.70 & 0.46 & 3.23 & 0.50 & 0.71 & 0.71 & 1.32 & 1.32 & 2.72 & 0.46 & 5.59 \\ 
82 & VCC 1948 & -16.15 & 1.02 & 0.34 & 0.94 & 0.39 & 0.52 & 0.56 & 1.12 & 1.15 & 1.00 & 0.36 & 0.86 \\ 
  &  &  & 1.05 & 0.42 & 0.99 & 0.47 & 0.59 & 0.67 & 1.25 & 1.27 & 1.04 & 0.43 & 0.64 \\ 
83 & VCC 1627 & -15.97 & 2.02 & 0.52 & 1.98 & 0.57 & 1.85 & 0.79 & 1.92 & 1.41 & 1.92 & 0.65 & 3.51 \\ 
  &  &  & 1.99 & 0.81 & 2.00 & 0.89 & 1.92 & 1.17 & 1.87 & 1.80 & 1.98 & 1.03 & 2.42 \\ 
84 & VCC 1440 & -15.95 & 2.13 & 1.11 & 2.14 & 1.20 & 1.55 & 1.51 & 2.09 & 2.09 & 2.14 & 1.24 & 2.35 \\ 
  &  &  & 2.15 & 1.15 & 2.17 & 1.25 & 1.56 & 1.56 & 2.13 & 2.13 & 2.18 & 1.29 & 2.33 \\ 
85 & VCC 230 & -16.18 & 2.90 & 0.34 & 2.94 & 0.39 & 2.03 & 0.57 & 1.16 & 1.15 & 2.94 & 0.39 & 6.40 \\ 
  &  &  & 2.92 & 0.38 & 2.98 & 0.42 & 1.79 & 0.61 & 1.21 & 1.20 & 2.99 & 0.43 & 6.32 \\ 
86 & VCC 2050 & -15.90 & 2.12 & 0.40 & 2.31 & 0.44 & 0.71 & 0.64 & 1.24 & 1.24 & 2.25 & 0.44 & 4.24 \\ 
  &  &  & 2.10 & 0.46 & 2.30 & 0.50 & 0.71 & 0.71 & 1.32 & 1.32 & 2.20 & 0.49 & 4.05 \\ 
87 & VCC 1993 & -15.91 & 1.23 & 0.58 & 1.22 & 0.64 & 0.87 & 0.87 & 1.50 & 1.50 & 1.21 & 0.65 & 1.47 \\ 
  &  &  & 1.21 & 0.70 & 1.15 & 0.77 & 1.03 & 1.03 & 1.67 & 1.67 & 1.17 & 0.75 & 1.21 \\ 
88 & VCC 751 & -15.83 & 2.20 & 0.74 & 2.43 & 0.81 & 1.10 & 1.08 & 1.72 & 1.72 & 2.42 & 0.81 & 3.49 \\ 
  &  &  & 2.37 & 0.89 & 4.39 & 0.97 & 1.26 & 1.26 & 1.88 & 1.88 & 2.71 & 0.93 & 3.48 \\ 
89 & VCC 1828 & -15.97 & 2.48 & 0.48 & 2.68 & 0.53 & 0.78 & 0.74 & 1.36 & 1.36 & 2.51 & 0.50 & 4.99 \\ 
  &  &  & 2.43 & 0.58 & 2.73 & 0.63 & 0.87 & 0.87 & 1.50 & 1.50 & 2.45 & 0.59 & 4.61 \\ 
90 & VCC 538 & -16.50 & 2.39 & 0.65 & 2.41 & 0.72 & 0.97 & 0.97 & 1.60 & 1.60 & 2.71 & 0.81 & 4.29 \\ 
  &  &  & 2.46 & 0.67 & 2.49 & 0.73 & 0.99 & 0.99 & 1.62 & 1.62 & 3.00 & 0.82 & 4.42 \\ 
91 & VCC 1407 & -15.78 & 2.16 & 0.44 & 2.18 & 0.49 & 2.45 & 0.70 & 1.31 & 1.31 & 2.18 & 0.49 & 4.10 \\ 
  &  &  & 2.12 & 0.51 & 2.15 & 0.57 & 1.97 & 0.79 & 1.41 & 1.41 & 2.14 & 0.55 & 3.86 \\ 
92 & VCC 1886 & -15.64 & 2.74 & 0.08 & 2.98 & 0.10 & 0.37 & 0.19 & 0.61 & 0.61 & 2.77 & 0.08 & 6.90 \\ 
  &  &  & 2.69 & 0.10 & 2.95 & 0.12 & 0.42 & 0.23 & 0.67 & 0.67 & 2.71 & 0.10 & 6.71 \\ 
93 & VCC 1199 & -15.69 & 2.09 & 1.05 & 2.09 & 1.14 & 2.10 & 1.45 & 2.04 & 2.04 & 2.10 & 1.32 & 2.00 \\ 
  &  &  & 2.10 & 1.08 & 2.09 & 1.17 & 2.10 & 1.47 & 2.06 & 2.06 & 2.09 & 1.34 & 2.00 \\ 
94 & VCC 1743 & -15.82 & 1.25 & 0.29 & 1.25 & 0.33 & 0.52 & 0.49 & 1.05 & 1.05 & 1.25 & 0.31 & 2.14 \\ 
  &  &  & 1.34 & 0.31 & 1.39 & 0.35 & 0.52 & 0.52 & 1.09 & 1.09 & 1.36 & 0.33 & 2.43 \\ 
95 & VCC 1539 & -15.60 & 2.28 & 0.54 & 2.30 & 0.59 & 1.76 & 0.82 & 1.47 & 1.45 & 2.27 & 0.52 & 4.26 \\ 
  &  &  & 2.26 & 0.44 & 2.28 & 0.49 & 1.78 & 0.69 & 1.35 & 1.31 & 2.26 & 0.44 & 4.49 \\ 
96 & VCC 1185 & -15.56 & 2.70 & 0.47 & 2.82 & 0.52 & 1.25 & 0.73 & 1.35 & 1.35 & 2.71 & 0.48 & 5.57 \\ 
  &  &  & 2.75 & 0.56 & 3.00 & 0.62 & 0.98 & 0.85 & 1.48 & 1.48 & 2.74 & 0.56 & 5.41 \\ 
97 & VCC 1826 & -15.43 & 3.10 & 0.48 & 3.15 & 0.53 & 1.54 & 0.74 & 1.36 & 1.35 & 3.15 & 0.56 & 6.33 \\ 
  &  &  & 3.10 & 0.49 & 3.16 & 0.54 & 1.48 & 0.76 & 1.38 & 1.38 & 3.19 & 0.57 & 6.26 \\ 
98 & VCC 1512 & -15.80 & \nodata & \nodata & \nodata & \nodata & \nodata & \nodata & \nodata & \nodata & \nodata & \nodata & \nodata \\ 
  &   &   & \nodata & \nodata & \nodata & \nodata & \nodata & \nodata & \nodata & \nodata & \nodata & \nodata & \nodata \\ 
99 & VCC 1489 & -15.56 & 2.56 & 0.19 & 2.72 & 0.22 & 1.26 & 0.36 & 0.87 & 0.87 & 2.68 & 0.22 & 6.12 \\ 
  &  &  & 2.41 & 0.22 & 2.57 & 0.25 & 0.72 & 0.40 & 0.92 & 0.92 & 2.50 & 0.24 & 5.67 \\ 
100 & VCC 1661 & -15.12 & 2.72 & 0.54 & 2.80 & 0.60 & 1.28 & 0.83 & 1.45 & 1.45 & 2.71 & 0.52 & 5.41 \\ 
  &  &  & 2.72 & 0.53 & 2.84 & 0.59 & 1.26 & 0.81 & 1.44 & 1.44 & 2.70 & 0.50 & 5.44 \\ 
\enddata 
\tablecomments{For each galaxy, the first row indicates the $g$-band values and the second the $z$-band values. Cols.~(1)~and~(2) list, respectively, the ACSVCS identification number of each galaxy \citep{Cot2004} and the Virgo Cluster Catalogue number \citep{Bin1985}. Col.~(3) lists the absolute $B$-band magnitudes of the galaxies, computed using apparent magnitudes from \citet{Bin1985}, extinctions from \citet{Sch1998} (for the Landolt $B$ filter), and distances from \citet{Bla2009}. Cols.~(4)-(8) are $\gamma_{\rm 3D}=-d\log{j(r)}/d\log{r}$ (where $j(r)$ is in $\mathcal{L}_{\odot}$/pc$^3$ and $r$ is in parsecs) at $0.5\%$, $1\%$, $5\%$, and $30\%$ of each galaxy's effective radius $R_e$, and at $0\Sec1$ (for comparison with \citetalias{Geb1996} and \citetalias{Lau2007}). The left hand side of each column gives $\gamma_{\rm 3D}$ for the total profile, whereas the right side is obtained from the deprojection of the S\'{e}rsic component only, for those galaxies that are nucleated. Col.~(10) denotes $\Delta_{\rm 3D}$, which quantifies the extent to which a light deficit ($\Delta_{\rm 3D}<0$) or excess ($\Delta_{\rm 3D}>0$) exists in the inner region of a given galaxy.  Note that we were not able to fit surface brightness profiles to a handful of ACSVCS galaxies due to the presence of dust and were thus unable to extract values for $\gamma_{\rm 3D}$ and $\Delta_{\rm 3D}$ for these.} 
\end{deluxetable} 
\clearpage

\clearpage 
\begin{deluxetable}{clcccccccccccc} 
\tabletypesize{\scriptsize} 
\tablecolumns{14}  
\tablecaption{$\gamma_{\rm 3D}$ and $\Delta_{\rm 3D}$ values computed for ACS Fornax Cluster Survey galaxies \label{tab:acsfcs}} 
\tablewidth{0pt} 
\tablehead{ 
\colhead{ID} & \colhead{Name} & \colhead{$M_B$} & \multicolumn{2}{c}{$\gamma_{\rm 3D}(0.5\%R_e)$} & \multicolumn{2}{c}{$\gamma_{\rm 3D}(1\%R_e)$} & \multicolumn{2}{c}{$\gamma_{\rm 3D}(5\%R_e)$} & \multicolumn{2}{c}{$\gamma_{\rm 3D}(30\%R_e)$} & \multicolumn{2}{c}{$\gamma_{\rm 3D}(0\Sec1)$} & \colhead{$\Delta_{\rm 3D}$} \\ 
\colhead{(1)} & \colhead{(2)} & \colhead{(3)} & \multicolumn{2}{c}{(4)} & \multicolumn{2}{c}{(5)} & \multicolumn{2}{c}{(6)} & \multicolumn{2}{c}{(7)} & \multicolumn{2}{c}{(8)} & \colhead{(9)} \\ 
} 
\startdata 
1 & FCC 21 & -22.30 & 1.73 & \nodata & 1.83 & \nodata & 2.10 & \nodata & 2.51 & \nodata & 1.30 & \nodata & -0.46 \\ 
  &  &  & 1.75 & \nodata & 1.85 & \nodata & 2.12 & \nodata & 2.52 & \nodata & 1.25 & \nodata & -0.64 \\ 
2 & FCC 213 & -21.05 & 0.57 & \nodata & 0.54 & \nodata & 2.13 & \nodata & 2.58 & \nodata & 1.01 & \nodata & -2.10 \\ 
  &  &  & 0.58 & \nodata & 0.48 & \nodata & 2.11 & \nodata & 2.58 & \nodata & 1.00 & \nodata & -2.14 \\ 
3 & FCC 219 & -20.67 & 0.98 & 1.01 & 0.67 & 1.10 & 1.41 & 1.41 & 2.00 & 2.00 & 0.98 & 1.01 & 0.29 \\ 
  &  &  & 0.98 & 1.01 & 0.70 & 1.10 & 1.41 & 1.41 & 2.01 & 2.01 & 0.98 & 1.01 & 0.27 \\ 
4 & NGC 1340 & -20.38 & 1.36 & 1.24 & 1.25 & 1.34 & 1.65 & 1.65 & 2.20 & 2.20 & 1.41 & 1.18 & 0.78 \\ 
  &  &  & 1.33 & 1.30 & 1.36 & 1.40 & 1.71 & 1.71 & 2.24 & 2.24 & 1.40 & 1.23 & 0.65 \\ 
5 & FCC 167 & -20.41 & \nodata & \nodata & \nodata & \nodata & \nodata & \nodata & \nodata & \nodata & \nodata & \nodata & \nodata \\ 
  &   &   & \nodata & \nodata & \nodata & \nodata & \nodata & \nodata & \nodata & \nodata & \nodata & \nodata & \nodata \\ 
6 & FCC 276 & -19.71 & 1.72 & 1.68 & 1.63 & 1.78 & 2.07 & 2.07 & 2.48 & 2.48 & 1.74 & 1.61 & 0.38 \\ 
  &  &  & 1.71 & 1.71 & 1.56 & 1.81 & 2.09 & 2.09 & 2.49 & 2.49 & 1.73 & 1.63 & 0.38 \\ 
7 & FCC 147 & -19.62 & 1.61 & 1.53 & 1.63 & 1.63 & 1.93 & 1.93 & 2.39 & 2.39 & 1.65 & 1.48 & 0.86 \\ 
  &  &  & 1.56 & 1.56 & 1.66 & 1.66 & 1.96 & 1.96 & 2.41 & 2.41 & 1.53 & 1.52 & 0.78 \\ 
8 & IC 2006 & -19.37 & 2.06 & 0.99 & 2.07 & 1.08 & 1.63 & 1.38 & 1.99 & 1.98 & 2.07 & 1.02 & 2.68 \\ 
  &  &  & 2.12 & 1.02 & 2.13 & 1.11 & 1.68 & 1.41 & 2.01 & 2.01 & 2.12 & 1.05 & 2.75 \\ 
9 & FCC 83 & -19.18 & 1.65 & 1.40 & 1.57 & 1.50 & 1.81 & 1.81 & 2.31 & 2.31 & 1.65 & 1.40 & 0.96 \\ 
  &  &  & 1.68 & 1.51 & 1.63 & 1.62 & 1.92 & 1.92 & 2.38 & 2.38 & 1.69 & 1.50 & 0.64 \\ 
10 & FCC 184 & -19.19 & 1.24 & 0.92 & 1.31 & 1.01 & 1.56 & 1.30 & 2.16 & 1.92 & 1.28 & 0.97 & 0.37 \\ 
  &  &  & 1.51 & 0.90 & 1.49 & 0.99 & 1.52 & 1.28 & 1.93 & 1.90 & 1.49 & 1.01 & 1.41 \\ 
11 & FCC 63 & -18.83 & 1.73 & 1.45 & 1.70 & 1.55 & 1.86 & 1.86 & 2.34 & 2.34 & 1.72 & 1.50 & 0.89 \\ 
  &  &  & 1.68 & 1.55 & 1.62 & 1.66 & 1.95 & 1.95 & 2.41 & 2.41 & 1.66 & 1.61 & 0.52 \\ 
12 & FCC 193 & -18.87 & 1.97 & 0.98 & 2.00 & 1.07 & 1.57 & 1.37 & 1.98 & 1.98 & 1.99 & 1.04 & 2.44 \\ 
  &  &  & 1.95 & 1.06 & 1.96 & 1.15 & 1.54 & 1.46 & 2.05 & 2.05 & 1.96 & 1.11 & 2.17 \\ 
13 & FCC 170 & -18.76 & 1.90 & 0.69 & 1.91 & 0.75 & 1.86 & 1.01 & 1.68 & 1.65 & 1.92 & 0.80 & 2.77 \\ 
  &  &  & 1.85 & 0.69 & 1.86 & 0.76 & 1.81 & 1.02 & 1.68 & 1.65 & 1.87 & 0.81 & 2.59 \\ 
14 & FCC 153 & -18.65 & 1.75 & 0.77 & 1.75 & 0.84 & 1.80 & 1.11 & 1.75 & 1.75 & 1.75 & 0.84 & 2.13 \\ 
  &  &  & 1.65 & 0.77 & 1.65 & 0.84 & 1.61 & 1.11 & 1.75 & 1.75 & 1.65 & 0.84 & 1.70 \\ 
15 & FCC 177 & -18.37 & 2.66 & 0.44 & 2.74 & 0.48 & 1.46 & 0.67 & 1.93 & 1.24 & 2.68 & 0.45 & 6.40 \\ 
  &  &  & 2.63 & 0.44 & 2.73 & 0.48 & 1.36 & 0.67 & 1.79 & 1.23 & 2.67 & 0.46 & 6.15 \\ 
16 & FCC 47 & -18.06 & 2.10 & 1.50 & 2.10 & 1.60 & 2.17 & 1.90 & 2.47 & 2.37 & 2.11 & 1.44 & 2.28 \\ 
  &  &  & 2.14 & 1.49 & 2.13 & 1.59 & 2.19 & 1.89 & 2.48 & 2.37 & 2.15 & 1.43 & 2.49 \\ 
17 & FCC 43 & -18.04 & 3.96 & 0.80 & 0.87 & 0.87 & 1.15 & 1.15 & 1.78 & 1.78 & 4.08 & 0.78 & 4.41 \\ 
  &  &  & 3.85 & 0.79 & 0.86 & 0.86 & 1.14 & 1.14 & 1.77 & 1.77 & 3.99 & 0.77 & 4.25 \\ 
18 & FCC 190 & -18.11 & 1.98 & 0.69 & 2.02 & 0.76 & 1.03 & 1.02 & 1.65 & 1.65 & 1.99 & 0.71 & 3.21 \\ 
  &  &  & 1.89 & 0.75 & 1.94 & 0.82 & 1.10 & 1.10 & 1.73 & 1.73 & 1.90 & 0.77 & 2.86 \\ 
19 & FCC 310 & -18.04 & 1.77 & 0.79 & 1.73 & 0.86 & 1.65 & 1.14 & 1.96 & 1.77 & 1.78 & 0.76 & 2.83 \\ 
  &  &  & 1.83 & 0.77 & 1.78 & 0.84 & 1.68 & 1.12 & 1.96 & 1.75 & 1.84 & 0.74 & 3.07 \\ 
20 & FCC 249 & -18.25 & 1.79 & 1.12 & 1.79 & 1.21 & 1.52 & 1.52 & 2.10 & 2.10 & 1.82 & 1.25 & 1.72 \\ 
  &  &  & 1.84 & 1.16 & 2.05 & 1.26 & 1.57 & 1.57 & 2.14 & 2.14 & 1.93 & 1.31 & 1.81 \\ 
21 & FCC 148 & -17.96 & 2.49 & 0.87 & 2.42 & 0.96 & 2.05 & 1.25 & 2.02 & 1.87 & 2.46 & 0.91 & 4.31 \\ 
  &  &  & 2.42 & 0.87 & 2.37 & 0.96 & 1.95 & 1.25 & 1.93 & 1.87 & 2.40 & 0.91 & 3.94 \\ 
22 & FCC 255 & -17.83 & 3.17 & 0.47 & 2.69 & 0.53 & 0.89 & 0.75 & 1.45 & 1.40 & 2.98 & 0.50 & 6.80 \\ 
  &  &  & 3.14 & 0.44 & 2.67 & 0.50 & 0.89 & 0.71 & 1.45 & 1.38 & 2.95 & 0.48 & 6.78 \\ 
23 & FCC 277 & -17.82 & 1.98 & 0.60 & 2.06 & 0.66 & 1.15 & 0.91 & 1.54 & 1.54 & 2.09 & 0.67 & 3.42 \\ 
  &  &  & 1.98 & 0.65 & 2.06 & 0.72 & 1.11 & 0.97 & 1.61 & 1.61 & 2.08 & 0.73 & 3.29 \\ 
24 & FCC 55 & -17.74 & 2.51 & 0.39 & 2.61 & 0.43 & 1.38 & 0.62 & 1.22 & 1.22 & 2.60 & 0.43 & 5.42 \\ 
  &  &  & 2.67 & 0.46 & 2.77 & 0.51 & 1.06 & 0.72 & 1.34 & 1.34 & 2.75 & 0.50 & 5.57 \\ 
25 & FCC 152 & -17.30 & -0.02 & \nodata & -0.03 & \nodata & 0.42 & \nodata & 1.03 & \nodata & -0.03 & \nodata & -0.60 \\ 
  &  &  & -0.03 & \nodata & -0.03 & \nodata & 0.57 & \nodata & 1.21 & \nodata & -0.04 & \nodata & -0.82 \\ 
26 & FCC 301 & -17.31 & 2.44 & 0.39 & 2.48 & 0.43 & 0.69 & 0.62 & 1.22 & 1.22 & 2.97 & 0.50 & 5.07 \\ 
  &  &  & 2.36 & 0.38 & 2.40 & 0.43 & 0.99 & 0.62 & 1.21 & 1.21 & 3.00 & 0.50 & 4.87 \\ 
27 & FCC 335 & -17.30 & 2.41 & 0.15 & 2.43 & 0.18 & 4.62 & 0.31 & 0.79 & 0.79 & 2.44 & 0.19 & 5.80 \\ 
  &  &  & 2.56 & 0.16 & 2.59 & 0.19 & 1.45 & 0.31 & 0.80 & 0.80 & 2.60 & 0.19 & 6.22 \\ 
28 & FCC 143 & -17.19 & 1.75 & 1.30 & 1.74 & 1.40 & 1.71 & 1.71 & 2.24 & 2.24 & 1.74 & 1.42 & 1.13 \\ 
  &  &  & 1.78 & 1.29 & 1.77 & 1.39 & 1.70 & 1.70 & 2.23 & 2.23 & 1.77 & 1.43 & 1.18 \\ 
29 & FCC 95 & -16.94 & 2.55 & 0.69 & 3.38 & 0.76 & 1.02 & 1.02 & 1.65 & 1.65 & 2.85 & 0.74 & 4.61 \\ 
  &  &  & 5.46 & 0.81 & 0.89 & 0.88 & 1.16 & 1.16 & 1.79 & 1.79 & 2.05 & 0.84 & 5.42 \\ 
30 & FCC 136 & -16.66 & 2.68 & 0.68 & 2.82 & 0.75 & 1.17 & 1.01 & 1.64 & 1.64 & 2.71 & 0.70 & 4.97 \\ 
  &  &  & 2.78 & 0.79 & 2.76 & 0.86 & 1.20 & 1.14 & 1.77 & 1.77 & 2.79 & 0.79 & 4.86 \\ 
31 & FCC 182 & -16.61 & 2.31 & 0.72 & 2.55 & 0.79 & 1.08 & 1.05 & 1.69 & 1.69 & 2.44 & 0.81 & 3.68 \\ 
  &  &  & 2.01 & 0.69 & 2.51 & 0.76 & 1.02 & 1.02 & 1.65 & 1.65 & 4.23 & 0.80 & 3.19 \\ 
32 & FCC 204 & -16.76 & 2.60 & 0.11 & 2.71 & 0.14 & 1.36 & 0.25 & 1.65 & 0.69 & 2.67 & 0.13 & 7.35 \\ 
  &  &  & 2.64 & 0.26 & 2.77 & 0.30 & 1.26 & 0.45 & 1.65 & 0.99 & 2.73 & 0.28 & 6.72 \\ 
33 & FCC 119 & -16.60 & -0.03 & \nodata & -0.03 & \nodata & 0.46 & \nodata & 1.06 & \nodata & -0.04 & \nodata & -0.64 \\ 
  &  &  & -0.03 & \nodata & -0.02 & \nodata & 0.55 & \nodata & 1.18 & \nodata & -0.04 & \nodata & -0.78 \\ 
34 & FCC 90 & -16.50 & 1.91 & 0.78 & 1.95 & 0.86 & 1.14 & 1.14 & 1.77 & 1.77 & 1.98 & 0.88 & 2.74 \\ 
  &  &  & 2.06 & 0.74 & 2.14 & 0.81 & 1.08 & 1.08 & 1.71 & 1.71 & 2.15 & 0.82 & 3.27 \\ 
35 & FCC 26 & -16.56 & -0.02 & \nodata & -0.04 & \nodata & 0.49 & \nodata & 1.36 & \nodata & -0.05 & \nodata & -1.11 \\ 
  &  &  & -0.02 & \nodata & -0.04 & \nodata & 0.51 & \nodata & 1.25 & \nodata & -0.04 & \nodata & -0.93 \\ 
36 & FCC 106 & -16.44 & 2.40 & 0.68 & 2.52 & 0.74 & 1.07 & 1.00 & 1.64 & 1.64 & 2.64 & 0.77 & 4.30 \\ 
  &  &  & 2.44 & 0.74 & 2.59 & 0.81 & 1.09 & 1.08 & 1.72 & 1.72 & 2.75 & 0.84 & 4.22 \\ 
37 & FCC 19 & -16.42 & 2.75 & 0.26 & 2.82 & 0.30 & 0.95 & 0.45 & 1.00 & 1.00 & 2.80 & 0.29 & 6.47 \\ 
  &  &  & 2.82 & 0.36 & 3.07 & 0.40 & 0.60 & 0.58 & 1.17 & 1.17 & 2.89 & 0.38 & 6.40 \\ 
38 & FCC 202 & -16.27 & 2.59 & 0.45 & 2.70 & 0.50 & 1.07 & 0.71 & 1.32 & 1.32 & 2.68 & 0.50 & 5.44 \\ 
  &  &  & 2.56 & 0.40 & 2.65 & 0.45 & 1.15 & 0.65 & 1.24 & 1.24 & 2.70 & 0.46 & 5.48 \\ 
39 & FCC 324 & -16.25 & 2.79 & 0.26 & 3.34 & 0.30 & 0.54 & 0.46 & 1.01 & 1.01 & 2.96 & 0.28 & 6.36 \\ 
  &  &  & 2.73 & 0.28 & 7.05 & 0.32 & 0.48 & 0.48 & 1.04 & 1.04 & 2.91 & 0.29 & 6.19 \\ 
40 & FCC 288 & -16.30 & 2.31 & 0.25 & 2.34 & 0.29 & 0.50 & 0.44 & 0.98 & 0.98 & 2.35 & 0.29 & 5.34 \\ 
  &  &  & 2.30 & 0.21 & 2.35 & 0.25 & 0.77 & 0.39 & 0.91 & 0.91 & 2.37 & 0.25 & 5.48 \\ 
41 & FCC 303 & -16.02 & 2.72 & 0.35 & 2.75 & 0.39 & 2.68 & 0.57 & 1.16 & 1.15 & 2.73 & 0.37 & 6.03 \\ 
  &  &  & 2.72 & 0.48 & 2.78 & 0.53 & 1.88 & 0.74 & 1.36 & 1.36 & 2.72 & 0.49 & 5.70 \\ 
42 & FCC 203 & -16.13 & 2.56 & 0.37 & 2.76 & 0.42 & 0.71 & 0.60 & 1.19 & 1.19 & 2.66 & 0.40 & 5.65 \\ 
  &  &  & 2.62 & 0.42 & 3.08 & 0.47 & 0.71 & 0.67 & 1.27 & 1.27 & 2.82 & 0.45 & 5.62 \\ 
43 & FCC 100 & -16.13 & 2.65 & 0.32 & 2.72 & 0.36 & 0.72 & 0.53 & 1.11 & 1.11 & 2.66 & 0.33 & 6.08 \\ 
  &  &  & 2.62 & 0.28 & 2.70 & 0.32 & 0.76 & 0.48 & 1.04 & 1.04 & 2.64 & 0.29 & 6.15 \\ 
\enddata 
\tablecomments{For each galaxy, the first row indicates the $g$-band values and the second the $z$-band values. Cols.~(1)~and~(2) list, respectively, the ACSFCS identification number of each galaxy \citep{Cot2004} and the Fornax Cluster Catalogue number \citep{Fer1989}. Col.~(3) lists the absolute $B$-band magnitudes of the galaxies, computed using apparent magnitudes from NED, extinctions from \citet{Sch1998} (for the Landolt $B$ filter), and distances from \citet{Bla2009}. Cols.~(4)-(8) are $\gamma_{\rm 3D}=-d\log{j(r)}/d\log{r}$ (where $j(r)$ is in $\mathcal{L}_{\odot}$/pc$^3$ and $r$ is in parsecs) at $0.5\%$, $1\%$, $5\%$, and $30\%$ of each galaxy's effective radius $R_e$, and at $0\Sec1$ (for comparison with \citetalias{Geb1996} and \citetalias{Lau2007}). The left hand side of each column gives $\gamma_{\rm 3D}$ for the total profile, whereas the right side is obtained from the profile deprojected without the fit to the nucleus, for those galaxies that are nucleated. Col.~(10) denotes $\Delta_{\rm 3D}$, which quantifies the extent to which a light deficit ($\Delta_{\rm 3D}<0$) or excess ($\Delta_{\rm 3D}>0$) exists in the inner region of a given galaxy.  Note that we were not able to fit surface brightness profiles to FCC~167 due to the presence of dust and were thus unable to extract values for $\gamma_{\rm 3D}$ and $\Delta_{\rm 3D}$ for this galaxy.} 
\end{deluxetable} 
\clearpage

\bibliographystyle{apj}
\bibliography{biblio}

\end{document}